\documentclass{aa}
\usepackage{txfonts,psfig,graphicx,natbib,aalongtable,colortbl}

\begin{document}
   \title{TNOs are Cool: A survey of the trans-Neptunian region. VIII. Combined Herschel PACS and SPIRE observations of 9 bright targets at 70--500 $\mu$m\thanks{Herschel is an ESA space observatory with science instruments provided by European led Principal Investigator consortia and with important participation from NASA.}
}
\author{S. Fornasier           \inst{1,2}\and
        E. Lellouch \inst{1}\and
        T. M\"uller \inst{3}\and
        P. Santos-Sanz \inst{1,4}\and
        P. Panuzzo \inst{5,6}\and
        C. Kiss \inst{7}\and
        T. Lim \inst{8}\and
        M. Mommert \inst{9}\and
         D. Bockel{\'e}e-Morvan\inst{1}\and
        E. Vilenius \inst{3}\and
          J. Stansberry \inst{10}\and
         G.P. Tozzi\inst{11}\and
        S. Mottola  \inst{9}\and
        A. Delsanti  \inst{1,12}\and
        J. Crovisier \inst{1}\and
         R. Duffard \inst{4}\and
        F. Henry \inst{1}\and
        P. Lacerda \inst{13}\and
        A. Barucci \inst{1}\and
         A. Gicquel \inst{1}
         }
 \offprints{Sonia Fornasier, LESIA-Obs. de Paris, 5 Place J. Janssen, 92195 Meudon Pricipal Cedex, France}
  \institute{LESIA, Observatoire de Paris, CNRS, UPMC Univ Paris 06, Univ. Paris-Diderot, 5 Place J. Janssen, 5 place Jules
Janssen, 92195 Meudon Pricipal Cedex, France
              \email{sonia.fornasier@obspm.fr}
         \and Univ Paris Diderot, Sorbonne Paris Cit\'{e}, 4 rue Elsa Morante, 75205 Paris Cedex 13, France
          \and  Max-Planck-Institut f\"ur  extraterrestrische Physik (MPE), Giessenbachstrasse,
85748 Garching, Germany
  \and  Instituto de Astrof\'isica de Andaluc\'ia (CSIC), Glorieta de la Astronom\'ia 
s/n, 18008 Granada, Spain
        \and  GEPI, Observatoire de Paris, CNRS, Univ. Paris Diderot, Place Jules Janssen, 92195 Meudon Pricipal Cedex, France
       \and CEA, Laboratoire AIM, Irfu/SAp, 91191 Gif-sur-Yvette, France
       \and  Konkoly Observatory, Research Centre for Astronomy and Earth Sciences, Hungarian Academy of Sciences, Konkoly Thege 15-17, H-1121 Budapest, Hungary
\and  Space Science and Technology Department, Science and
Technology Facilities Council, Rutherford Appleton Laboratory,
Harwell Science and Innovation Campus, Didcot, Oxon UK, OX11
0QX, UK
        \and  Institute of Planetary Research, DLR, Rutherfordstrasse 2, 12489 Berlin, Germany
       \and  The University of Arizona, Tucson AZ 85721, USA
      \and  INAF--Osservatorio Astrofisico di Arcetri, Largo E. Fermi 5, 50125, Firenze, Italy
         \and  Aix Marseille Universit\'e, CNRS, LAM  (Laboratoire d'Astrophysique de Marseille) UMR 7326, 38 rue Fr\'ed\'eric Joliot-Curie, 13388 Marseille,  France
       \and  Astrophysics Research Centre, Queen’s University Belfast, Belfast BT7 1NN,
UK}

 \date{Received February 2013, accepted for publication on 12 April 2013.}

\abstract{}{Transneptunian objects (TNOs) are bodies populating the Kuiper Belt and they are
 believed to retain the most pristine and least altered material of the solar system. The Herschel Open Time Key Program entitled ``TNOs are Cool: A survey of the trans-Neptunian region'' has been awarded 373 hours to investigate the albedo, size distribution and thermal properties of TNOs and Centaurs. Here we focus on the brightest targets observed by both the PACS and SPIRE multiband photometers: the dwarf planet Haumea, six TNOs (Huya, Orcus, Quaoar, Salacia, 2002 UX25, and 2002 TC302), and two Centaurs (Chiron and Chariklo).}{Flux densities are derived from PACS and SPIRE instruments using optimised data reduction methods. The spectral energy distribution
obtained with the Herschel PACS and SPIRE instruments over 6 bands (centred at 70, 100, 160, 250, 350, and 500 $\mu$m), with Spitzer-MIPS at 23.7 and 71.4 $\mu$m, and with WISE at 11.6 and 22.1 $\mu$m in the case of 10199 Chariklo, has been modelled with the NEATM thermal model in order to derive the albedo, diameter, and beaming factor. For the Centaurs Chiron and Chariklo and for the 1000 km sized Orcus and Quaoar, a thermophysical model was also run to better constrain their thermal properties.}{We derive the size, albedo, and thermal properties, including thermal inertia and surface emissivity, for the 9 TNOs and Centaurs. Several targets show a significant decrease in their spectral emissivity longwards of $\sim$300 $\mu$m and especially at 500 $\mu$m. Using our size estimations and the mass values available in the literature,  we also derive the bulk densities for the binaries Quaoar/Weywot (2.18$^{+0.43}_{-0.36}$ g/cm$^3$), Orcus/Vanth (1.53$^{+0.15}_{-0.13}$ g/cm$^3$), and Salacia/Actea (1.29$^{+0.29}_{-0.23}$ g/cm$^3$). Quaoar's density is similar to that of the other dwarf planets Pluto and Haumea, and its value implies high contents of refractory materials mixed with ices.}{}


\titlerunning{TNOs are Cool program: PACS and SPIRE results}
\authorrunning{}
\maketitle

\section{Introduction}

\begin{figure*}[t]
\centering
\includegraphics[width=18cm,angle=0]{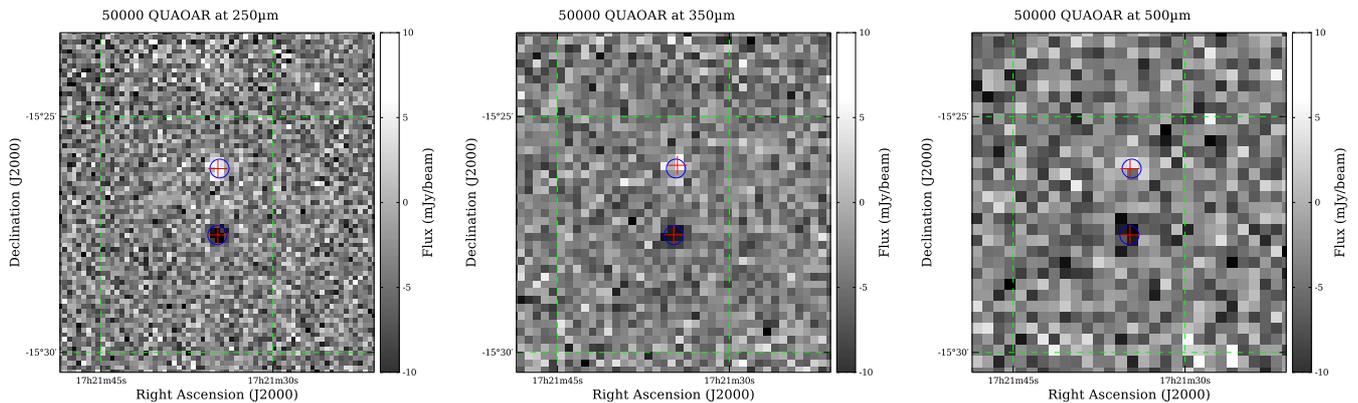}
\caption{Difference maps at 250, 350, and 500 $\mu$m of 50000 Quaoar. Circles show the expected positions of the target, according
to ephemerides, in the two epochs, while crosses show positions computed by the fitting routine.}
\label{fig_map_quaoar}
\end{figure*}

Transneptunian objects (TNOs), also known as Kuiper belt objects (KBOs), are thought to be frozen leftovers
from the formation period of the outer solar system. Investigation of their surface properties is essential
for understanding the formation and the evolution of the solar system, and sheds light on the composition of
the primordial protoplanetary disk.  \\
The Kuiper belt has been heavily perturbed dynamically, as indicated by the presence of bodies with highly inclined and/or very eccentric orbits and the existence of widely different dynamical classes (Gladman et al. 2008). Resonant objects are trapped in resonances with Neptune, with the majority of the discovered ones located in or near the 3:2  mean motion resonance, and called plutinos after their prototype, Pluto. Scattered disk objects (SDO) have high-eccentricity, high-inclination orbits and a perihelion distance near q=35 AU. Detached objects, previously called extended scattered disk objects, are located outside of interacting gravitational encounters with Neptune. Classicals are non-resonant and non-scattering objects with low eccentricity. In addition, the Centaurs are associated with the TNO population. Centaurs are closest to the Sun and have unstable orbits between those of Jupiter and Neptune. They cross the orbits of one or more giant planets. The Kuiper belt region seems to be the source of both short period comets and the Centaurs, which had been injected into their present orbit by gravitational instabilities and collisions (Tegler et al. 2008).

To date, more than 1600 TNOs and Centaurs have been discovered. Although these objects reside in more or less the same region of the solar system, they can have very different surface characteristics, and they reveal a richness of orbital and physical properties, with few apparent links between their orbital and surface properties (Doressoundiram et al. 2008).  TNOs and Centaurs cover a wide range of intrinsic colours from slightly bluish slopes (--10 \%/100 nm in
the reflectance) to the reddest gradients ($>$50\%/100 nm for Pholus) known in the solar system (Fornasier et al. 2004a, 2009; Doressoundiram et al. 2008). \\
Spectroscopy confirms the presence of ices of several kinds (Barucci et al. 2011).
Essentially, objects can be classified as water-ice rich (about 30 objects), volatile-ice rich (methane and perhaps nitrogen on bright objects such as Pluto, Eris, Makemake and Sedna) and featureless, although a couple of them show additional features, perhaps due to methanol or ammonia.

Knowledge of TNOs' albedos and sizes is important for constraining the surface composition and understand the dynamical evolution of the outer solar system. In particular, absolute albedos must be known to (i) properly model the spectra in terms of surface composition (ii) convert optical magnitudes into diameters, thereby deriving size distributions for the various sub-populations. Nevertheless, TNOs' albedos and sizes are among the most difficult to measure. Except for a few objects measured from direct imaging or stellar occultations, the combined measurements of the reflected light and thermal emission flux are required to derive TNOs' size and albedo. Except for a few objects measured from the ground, existing measurements of TNO thermal fluxes mostly come from Spitzer and more recently from Herschel space telescopes observations.
Based on Spitzer MIPS observations at 24 and 70 $\mu$m, the albedo and the size of 39 objects were published (Stansberry et al. 2008). These data indicate a large range in TNOs surface albedo, generally ranging from 3\% to 20\%, with a mean value of 8\%. Prominent exceptions are Eris, 2002 TX300, Makemake, and Haumea, which have albedos of 0.96$^{+0.09}_{-0.04}$ (Sicardy et al., 2011), 0.88$^{+0.15}_{-0.06}$ (Elliot et al., 2010), 0.7--0.9 (Lim et al. 2010), and 0.70-0.75 (Lellouch et al. 2010), respectively.

To better investigate the albedo, size distribution, and thermal properties of TNOs and Centaurs, an open time key programme was submitted for the Herschel Space Telescope, whose observational spectral range covers the thermal flux peaks of TNOs. This proposal, entitled {\it TNOs are Cool: A survey of the trans-Neptunian region} (M\"uller et al. 2009) has been awarded  372.7  hours  to perform radiometric
measurements of a large TNOs/Centaurs sample of about 130 objects, including 27 known multiple systems.
The 130 targets were selected on the basis of their predicted thermal flux in order to reach a suitable S/N with the PACS/SPIRE Herschel instruments. All the targets  have been observed with the PACS photometer instrument (3 bands centred at 70, 100, and
 160 $\mu$m), and only the brightest ones (11 objects) with the SPIRE instrument (with channels centred at 250, 350, and 500 $\mu$m).\\
The first results from this observing programme, obtained during the Herschel demonstration phase, were published by M\"uller et al. (2010), Lim et al. (2010), and Lellouch et al. (2010). Recently, the Herschel observations on 19 classicals Kuiper belt objects, 18 plutinos, and 15 scattered disk and detached objects were published by Vilenius et al. (2012), Mommert et al. (2012), and Santos-Sanz et al. (2012), respectively. The average geometric albedos of
the different dynamical classes are 17\% and 11\% for the cold and hot classicals, respectively (Vilenius et al. 2012), 8\% for the plutinos (Mommert et al. 2012), 7\% for the SDOs, and 17.0\% for the detached objects (excluding Eris), respectively (Santos-Sanz et al. 2012). \\
In this paper, we focus on the brightest objects that were observed both by PACS and SPIRE. We thus present results for two Centaurs (Chiron and Chariklo), the dwarf planet Haumea, two plutinos (Huya and Orcus), three classical TNOs (Quaoar, Salacia and 2002 UX25), and the 2:5 resonant object 2002 TC302.

\section{Observations and data reduction}

\begin{table*}[t]
\caption{Summary of Herschel observations}
{\footnotesize
\label{tabobs}
\begin{center}
\begin{tabular}{|l|l|l|c|c|c|c|c|} \hline
Target & Obs. ID & Date & UT start & Exp. time  & Instrum. & Bands & Dyn. class \\ 
       &         &      &          & (s)        &          & ($\mu$m) & \\ \hline
(2060) Chiron     & 1342195665  &2010-04-27  &23:51:55  & 1697 & SPIRE & 250/350/500  &  Centaur \\
                & 1342195675  & 2010-04-28 &19:26:34 & 1697  & SPIRE & 250/350/500  & \\
    &  1342195392--95 & 2010-04-28 &00:02:21 & 2272 &  PACS & 70/100/160 & \\
      & 1342195404--07 & 2010-04-28 & 07:05:37& 2272 &  PACS & 70/100/160 &  \\ \hline
(10199) Chariklo  & 1342202219 & 2010-08-05 & 13:09:34 & 2111 & SPIRE & 250/350/500  & Centaur \\
                & 1342202212 &  2010-08-04 & 17:37:11 & 2111 & SPIRE & 250/350/500  & \\ \
  & 1342202372--75 &  2010-08-10 & 17:51:05 & 1144 &  PACS & 70/100/160 & \\
  & 1342202570--73 &  2010-08-11 & 02:14:21 & 1144 &  PACS & 70/100/160 &  \\ \hline
(38628) Huya    & 1342201135 & 2010-07-22 & 12:35:04 & 2111 & SPIRE & 250/350/500  & Plutino \\
                & 1342201255 & 2010-07-26& 17:34:10 & 2111 & SPIRE & 250/350/500  &\\
    &1342202873--76 & 2010-08-11 & 19:01:34& 2272 &   PACS & 70/100/160 & \\
    &1342202914--17 & 2010-08-12 & 11:33:49& 2272 &   PACS & 70/100/160 &  \\ \hline
(50000) Quaoar &  1342205970--73 & 2010-10-06 & 23:13:43    & 2272 &   PACS & 70/100/160 &  Classical \\
               &  1342206017--20  & 2010-10-07 & 15:14:36 &       2272 &   PACS & 70/100/160 &  \\
     &  1342227654 &  2011-08-31 & 21:17:54 & 2388 & SPIRE & 250/350/500  & \\
               &  1342228351  & 2011-09-08 & 19:21:29 &  2388 & SPIRE & 250/350/500  & \\ \hline
(55637) 2002 UX25& 1342201146 & 2010-07-23 & 11:35:02 & 2111 & SPIRE & 250/350/500  & Classical \\
                 & 1342201324 & 2010-07-27 & 11:07:32& 2111 & SPIRE & 250/350/500  & \\
& 1342202881--84 & 2010-08-11 & 21:10:42 & 2272 &   PACS & 70/100/160 &  \\
  & 1342203035--38 & 2010-08-14 & 01:39:47& 2272 &  PACS & 70/100/160 &  \\ \hline
(84522) 2002 TC302 & 1342203092 &2010-08-16& 12:24:13 &2111 &  SPIRE & 250/350/500  & Res. 2:5\\
                   & 1342203285 & 2010-08-21& 12:29:38 &2111 &  SPIRE & 250/350/500  & \\
 & 1342214049--52 & 2011-02-10 & 15:37:12& 2272 &   PACS & 70/100/160 & \\
                   & 1342214159--62 & 2011-02-12 & 01:13:16& 2272 &   PACS & 70/100/160 &  \\ \hline
(90482) Orcus     & 1342187261  &2009-11-28  &23:52:02 & 1136 & SPIRE & 250/350/500  &  Plutino\\
            & 1342187262  &2009-11-29  &00:11:55 & 1136 & SPIRE & 250/350/500  & \\
            & 1342187522  &2009-12-01  &17:37:27 & 1136 & SPIRE & 250/350/500  & \\
            & 1342187523  &2009-12-01  &17:57:20 & 1136 & SPIRE & 250/350/500  & \\
    & 1342195997--6000 & 2010-05-08 & 23:56:35& 2272 &   PACS & 70/100/160 &  \\
                 & 1342196129--32 & 2010-05-10 & 06:37:57& 2272 &   PACS & 70/100/160 &  \\ \hline
(120347) Salacia & 1342198913--16 &  2010-06-22  & 00:56:50 & 2272 & PACS & 70/100/160 &  Classical \\
                 & 1342199133--36 &  2010-06-22  & 18:37:22 & 2272 & PACS & 70/100/160 &  \\
                  &1342236129 &     2012-01-01 & 13:51:00  & 2387 & SPIRE & 250/350/500  &\\
                 &1342236257 &  2012-01-03 & 06:43:35  & 2387 & SPIRE & 250/350/500  &\\ \hline
(136108)  Haumea & 1342212360 & 2011-01-07 & 07:09:24 &2111 &  SPIRE & 250/350/500  & Classical\\
                 & 1342212414 & 2011-01-09 & 09:05:52 &2111 &  SPIRE & 250/350/500  & \\
                 &  1342198851  &  2010-06-20 & 20:45:11 &  568    & PACS$^a$ & 100/160 &  \\
                 & 1342198903--06 & 2010-06-21 & 22:42:00 & 568    & PACS     & 70/100 & \\

\hline
\end{tabular}
\end{center}
}
\tablefoot{The columns represent the target name, observation identifier from the
Herschel Science Archive (OBSID), start-date/time (UT), duration, PACS/SPIRE photometer bands (in $\mu$m). $^a$: for Haumea, the PACS 100 and 160 $\mu$m fluxes are the mean values derived from the target lightcurve observations, which span 4.3 hours (the results on the Haumea lightcurve are presented in a separate paper by Santos-Sanz et al., in preparation).}
\end{table*}

\begin{table*}[t]
\caption{Colour corrected fluxes and geometric conditions for the observed TNOs and Centaurs.}  
{\footnotesize
\label{fluxHerschel}
\begin{center}
\begin{tabular}{|l|c|c|c|c|c|c|c|c|c|c|} \hline
Object &    F$_{70\mu m}$  & F$_{100\mu m}$ & F$_{160\mu m}$  & F$_{250\mu m}$  & F$_{350\mu m}$  & F$_{500\mu m}$  & r  & $\Delta$ & $\alpha$  & H$_{v}$ \\
       & (mJy) & (mJy) & (mJy) & (mJy) & (mJy) & (mJy) & (AU) & (AU) & ($^{o}$) & \\ \hline
(2060) Chiron   & 70.8$\pm$3.1 & 46.2$\pm$2.3 & 22.8$\pm$2.9 & 11.1$\pm$1.9 & 1$\sigma$=2.9 & 1$\sigma$=3.2  &    16.282  &  16.635   &  3.3 & 5.92$\pm$0.20$^{a}$ \\
(10199) Chariklo    &  142.9$\pm$5.9 & 99.1$\pm$4.4 & 50.8$\pm$4.3 & 26.0$\pm$2.4 &  12.8$\pm$1.9 & 1$\sigma$=3.0 & 13.861 & 13.750 & 4.2 & 7.40$\pm$0.25$^{a}$\\
(38628) Huya        &  47.6$\pm$1.7  & 43.2$\pm$1.9    &  25.6$\pm$2.4    & 14.8$\pm$1.9  & 9.0$\pm$1.8  & 1$\sigma$=3.2    &    28.665  &  28.767  &  2.0 & 5.04$\pm$0.03$^{b,c,d}$ \\
(50000) Quaoar & 32.1$\pm$2.4 & 41.5$\pm$3.1 & 29.9$\pm$2.4 & 17.3$\pm$1.9 & 10.4$\pm$1.8 & 7.3$\pm$2.5 & 43.136 &  43.355 & 1.2 & 2.73$\pm$0.06$^{b,c}$\\
(55637) 2002 UX25 &  25.8$\pm$1.3 & 26.0$\pm$1.5 & 20.2$\pm$3.8 & 11.6$\pm$2.8 & 1$\sigma$=3.1 & 1$\sigma$=3.3 & 41.609 &  41.334 & 1.4 & 3.87$\pm$0.02$^{b}$ \\
(84522) 2002 TC302 &  9.1$\pm$1.4  & 16.2$\pm$1.9 & 6.0$\pm$2.4 & 1$\sigma$=3.1 & 1$\sigma$=3.2 & 1$\sigma$=3.4 & 46.230 & 46.503 & 1.2 & 4.17$\pm$0.10$^{e,f}$ \\
(90482) Orcus  & 25.7$\pm$1.3 &  34.2$\pm$1.9   &  23.5$\pm$2.2   & 16.4$\pm$2.0  & 8.9$\pm$1.8 &  1$\sigma$=3.1   &   47.898  &  47.705  &  1.2 & 2.31$\pm$0.03$^{b,g}$ \\
(120347) Salacia & 30.0$\pm$1.2 & 37.8$\pm$2.0 & 28.1$\pm$2.7 & 11.0$\pm$1.7 & 7.3$\pm$1.6 & 1$\sigma$=2.8 & 44.213 & 44.281 & 1.3  & 4.25$\pm$0.05$^{h,i}$ \\
(136108) Haumea    & 17.3$\pm$3.4 & 23.5$\pm$0.8 & 22.2$\pm$1.4  & 16.3$\pm$2.0 & 10.7$\pm$1.8 & 1$\sigma$=2.8  &   50.971  &  50.961  &  1.1 & 0.43$\pm$0.01$^{b}$ \\ \hline \hline
\end{tabular}
\end{center}
}
\tablefoot{r and $\Delta$ are the target heliocentric and Herschel-centric distances, respectively, and $\alpha$ is the phase angle, calculated at the middle time of the PACS and SPIRE  observations.
F$_{\lambda}$ is the colour-corrected flux with its 1-$\sigma$ uncertainty. Entries such
as  1$\sigma$=2.9 indicate that the object was not detected, with the corresponding 1-$\sigma$
upper limit. For the H$_V$ magnitude values $^{a}$: see sections 4.1 and 5 of this paper; $^{b}$: Rabinowitz et al. 2007; $^{c}$: Romanishing \& Tegler, 2005; $^{d}$: Doressoundiram et al. 2005; $^{e}$: computed from Santos-Sanz et al. 2009 with linear coeff. =0.122 mag/$^o$; $^{f}$: computed from Sheppard 2010 with linear coeff. = 0.122 mag/$^o$; $^{g}$: Perna et al. 2010; $^{h}$: Benecchi et al. 2009; $^{i}$: Perna et al. 2013.}
\end{table*}

\subsection{Herschel-SPIRE observations}

The Spectral and Photometric Imaging REceiver (SPIRE) is the Herschel Space Observatory submillimetre camera and spectrometer, which contains a three-band imaging photometer operating at 250, 350, and 500 $\mu$m, and an imaging Fourier-transform spectrometer (FTS) which simultaneously covers its whole operating range of 194--671 $\mu$m (Griffin et al. 2010). For the observations of TNOs and Centaurs we used the imaging photometer that has a field of view of 4$\times$8$\arcmin$, observed simultaneously in all three spectral bands. All the targets were observed in small-map mode, except (90482) Orcus, which was observed in large map mode (see Lim et al. 2010) during the Herschel science demonstration phase.
In the small-map mode, the telescope was scanned across the sky at 30$\arcsec$/s, in two nearly orthogonal (at 84.8 deg) scan paths, covering uniformely an area of 5$\times$5$\arcmin$. The number of repeats varied between 12 and 15 for our targets. See Griffin et al. (2010) for additional information on the instruments' operation mode and performances.

The expected flux of our targets in the SPIRE bands is usually comparable to the SPIRE confusion noise, which is of 5.8, 6.3, and 6.8 mJy/beam at 250, 350, and 500 $\mu$m, respectively (Nguyen et al. 2010). To have reliable detections, for each target the SPIRE observations were performed at two epochs separated by a few hours/days, in order to have the target moving by a few beam sizes (see Table~\ref{tabobs} for the details on the observing conditions).
By subtracting the maps produced at the two epochs, we can obtain two images of the target and remove the background.

The processing of SPIRE data was executed up to level 1 (calibrated timelines) with the standard SPIRE pipeline (Swinyard et al., 2010), using the Herschel Interactive Processing Environment (HIPE\footnote{HIPE is a joint development by the Herschel Science Ground Segment Consortium, consisting of ESA, the NASA Herschel Science Center, and the HIFI, PACS, and SPIRE consortia members, see http://Herschel.esac.esa.int/DpHipeContributors.shtml} version 8.0), including the turnaround data (see the SPIRE pipeline documentation for more details).
The baseline in the signal for each bolometer was then removed by subtracting the median of the signal for each scan.
To remove any residual trend in signal timelines and minimize background differences between data
taken in the two epochs, we applied a destriping routine to baseline-subtracted data. The routine measured the difference between the signal registered by each bolometer for each scan and the signal on the
reconstructed map at the same sky coordinates. This difference, as a function of time along the scan
for each bolometer, was fitted with a polynomial of degree 5 and subtracted from the bolometer
signal timeline. A new map was reconstructed, preserving the flux, and the procedure was repeated for 20 iterations. \\
Maps were then produced for each epoch using the standard naive map making, projecting the data of each band on the same
World Coordinate System. Data can be affected by astrometry offsets between the two epochs due to pointing errors of the telescope, and, as a consequence, the confusion noise can leave residuals in the difference map. Thus we cross-correlated
the 250, 350, and 500 $\mu$m maps of the two epochs to measure astrometry offsets, which were then corrected in the data of the second epoch. Finally, the difference map was obtained by subtracting the two epoch maps pixel by pixel.
Figure~\ref{fig_map_quaoar} shows the difference maps in the three SPIRE bands for Quaoar.\\
In the difference map we obtain a positive and a negative image of the target. For each image, we fit the flux distribution in the pixels with a two-dimensional circular Gaussian with a fixed Full Width Half Maximum (FWHM), given by the SPIRE Observing Manual.
The fitting algorithm is implemented in HIPE and is based on the Levenberg-Marquardt method.
The position of the gaussian centre wass left as free parameter for the 250 and 350 $\mu$m bands, while for the 500 $\mu$m band, the centre was assumed to coincide with the one found by the 250 $\mu$m band Gaussian fit. \\
The fitting algorithm provides the uncertainty on the amplitude of the fitted Gaussians. Each pixel has flux errors that are converted into a weight of the pixel value within the algorithm. We computed the errors in two different ways.
In the first one, the errors of the difference map were calculated as the quadratic sum of errors in the two epochs maps.
Each epoch map has an associated error map that represents the standard deviation of data samples divided by the
square root of the number of data samples.
The second method consists in estimating the error as the standard deviation of fluxes measured in the
difference map around the sources. While the error with the first method is dominated by the instrumental noise,
the second one includes also the contribution of residual astrometry offsets between observations in the two epochs
and residual background variations. We found that the two methods give similar results, confirming that
confusion noise and instruments artefacts are properly removed. To be conservative, we finally used the greatest of the two error estimates for each source.

The amplitude of the fitted Gaussians represents the measured fluxes of the source at each epoch.
This flux must be divided by the pixelization factors (0.951, 0.931, and 0.902 for
250, 350, and 500 $\mu$m, respectively, for the default pixel sizes, see the SPIRE Observing Manual).
In fact, the SPIRE flux calibration is time-line based (Jy in beam). As a result, the signal level in a map pixel depends on how the square map pixel size compares to the size of the beam. Only in the limit of infinitely small map pixels would a pixel co-aligned with a point source register the true source flux density. For a given map pixel, the flux density value represents the average in-beam flux density measured by the detectors while pointing into that area. These pixelization factors are therefore needed to consider the effect of averaging time-line data in pixels when producing maps.
Pixelization also introduces a source of error of the order of 1.5\% of the flux (SPIRE Observing Manual), which we added in quadrature to uncertainties provided by the fitting algorithm. Finally, we also added 7\% of the
measured flux as the uncertainty on the absolute calibration (SPIRE Observing Manual).\\
Assuming that the flux of the source does not change between the two epochs, the final
flux is computed as the weighted
mean of the two flux estimates, and the associated uncertainty is the square root of the inverse of the sum of the weights.\\
Finally, the fluxes were colour-corrected. For both SPIRE and PACS instruments, fluxes are given as
the monochromatic flux densities assuming a power law as source spectrum across the band defined by the flux density at a standard frequency $\nu_0$ (corresponding to wavelengths of 250, 350, and 500 $\mu$m for SPIRE), and a spectral index $\alpha_S =-1$. The flux density of each target was then colour-corrected using the instrument bands profiles convolved with a black-body profile having the same temperature value (derived from a first NEATM modelling run, see Section 3)  of the object.  The final colour-corrected fluxes are given in Table~\ref{fluxHerschel}.

\subsection{Herschel-PACS observations}

The Photodetector Array Camera and Spectrometer (PACS) can be operated as photometer and spectrometer (Poglitsch et al., 2010). For our observing programme we used the PACS photometer, which has three bands centred at 70, 100, and 160 $\mu$m covering the 60--85 $\mu$m, 85--125 $\mu$m, and 125--210 $\mu$m range, respectively.
With two bolometer arrays, PACS simultaneously images two bands, the 70/160 $\mu$m or the 100/160 $\mu$m over a field of view of $\sim$1.75$\times$3.5$\arcmin$. \\
PACS observations of TNOs and Centaurs were performed in mini-scan map mode at two
epochs, separated by a time interval that corresponds to a motion of
$\sim$30$\arcsec$ of the target, allowing
for an optimal background subtraction (see Table\ref{tabobs} for the details on PACS and SPIRE observations). At each epoch
the target is observed in the ``blue'' (nominal wavelength of
70\,$\mu$m) and in the
``green'' (100\,$\mu$m) band twice, using two different scan position angles (red
channel - 160 $\mu$m -
data are always taken in parallel, either with the blue or
the green channel). This forms a series of four measurements in the blue and
green bands, and a series of eight measurements in the red band for a specific
target. The maps are taken in the medium scan speed (20\arcsec/sec) mode.
The observing strategy and sequences are described in Vilenius et al. (2012).

The PACS data were reduced using HIPE by means of adapted standard HIPE scripts. With HIPE we generated one {\it single} map per visit, filter, and scan direction.
To combine the images and to remove the background we
used two methods: {\it super-sky-subtracted} images (Santos-Sanz et al., 2012) and the  {\it double-differential} images (Mommert et al., 2012).
The {\it super-sky}  is constructed by masking the source (or an area surrounding the image centre when the target is too faint to be recognised in individual images)
in each individual image. Images were then combined to build a background map that is subtracted from each single map. Finally we co-add all the background-subtracted maps in the co-moving frame of the target. \\
The {\it double-differential} images were produced by combining the two visits maps of a target in a single map. This yields a positive and a negative beam of the moving source with background structures eliminated.  A duplicate of this image is shifted to match the positive beam of the original image with the negative one of the duplicate. By varying the proper motion vector between the two visits, the cross-correlation residuals for each trial vector is computed. Images are finally combined generating a double-differential image with one positive and two negative beams.\\
On the final maps we performed standard synthetic-aperture photometry on the localised source centroid. We measured the flux at the photocentre position for aperture radii ranging from 1 to 15 pixels, and then we applied an aperture correction technique (Howell 1989) for each aperture radius using the encircled energy fraction for a point source for the PACS instrument. From the aperture-corrected curve-of-growth we selected the optimum synthetic aperture to finally measure the target flux. The radius of the selected aperture is typically $\sim$ 1.0--1.25 times the point spread function FWHM (PSF FWHM in radius is 5.2$\arcsec$/7.7$\arcsec$/12$\arcsec$ in 70/100/160 $\mu$m bands, respectively). 

Uncertainties on the flux measurements were estimated by
means of a Monte-Carlo technique (Mueller et al., 2011), in which 200 artificial
sources, having the structure of the PSF for the specific band, are implanted on the background-subtracted final
maps in a square region of 50$\times$50$\arcsec$ around the target
photocentre (avoiding the region just surrounding the target but staying at the same time in a region with high coverage). Uncertainties were
computed as the standard deviation of these 200 fluxes, and finally
multiplied by a factor $\sqrt{2}$ , because the remaining
background was measured only once in the immediate vicinity
of the real target but two times in the rest of the image. \\
Overall, the data reduction procedure is exactly the same as adopted by Santos-Sanz et al. (2012) for the analyses of 15 scattered disk and detached TNOs within the same observing programme, so we refer to this paper for a detailed description of the PACS data reduction strategy. \\
The final colour-corrected fluxes are given in Table~\ref{fluxHerschel}.

\begin{table*}[t]
\caption{Spitzer-MIPS colour corrected fluxes at 23.68 and 71.42 $\mu$m, and, for Chariklo, WISE colour corrected fluxes at 11.6 and 22.1 $\mu$m. For 2002 TC302, the 24 $\mu$m flux is unreliable since the object is merged with a background source.}
{\footnotesize
\label{fluxSpitzer}
\begin{center}
\begin{tabular}{|l|l|c|c|c|c|c|c|c|} \hline
Object &  Obs. start     &F$_{11.6 \mu m}$ [mJy] & F$_{22.1 \mu m}$ [mJy] &F$_{24\mu m}$ [mJy] & F$_{71\mu m}$ [mJy] & r$_h$ (AU) & $\Delta$ (AU) & $\alpha$ ($^{o}$)\\  \hline
(2060) Chiron   & 2005-May-16 01:16:33     & &  & 54.44$\pm$1.64$^{1}$  &147.50$\pm$10.70$^{1}$ &13.462 & 13.238 & 4.2  \\
(10199) Chariklo    & 2006-Feb-15 15:13:04  & & & 61.02$\pm$1.86$^{1}$ & 179.41$\pm$11.50$^{1}$ & 13.164 & 12.890 & 4.2\\
(10199) Chariklo    & 2010-Feb-16 19:39:59  & 0.61$\pm$0.17 & 39.57$\pm$5.12 &  &  & 13.760 & 13.716 & 4.1\\
(38628) Huya        &  2004-Jan-27 09:32:09& &  & 3.45$\pm$0.11$^{1}$ & 55.03$\pm$4.20$^{1}$ & 29.326 & 29.250 & 2.0\\
(50000) Quaoar &  2006-Apr-03 23:09:45    & &   & 0.22$\pm$0.02$^{2,3}$ & 28.96$\pm$3.67$^{2,3}$ & 43.311 & 43.086 & 1.3 \\
(50000) Quaoar &  2005-Apr-07 09:17:49     & &  & 0.26$\pm$0.06$^{2,3}$ & 26.99$\pm$4.90$^{2,3}$ & 43.345 & 42.974 & 1.2 \\
(55637) 2002 UX25 & 2005-Jan-26 18:56:15   & &  & 0.44$\pm$0.04$^{1,2}$ &23.51$\pm$4.43$^{1,2}$& 42.369 & 42.382 & 1.4 \\
(84522) 2002 TC302 & 2005-Jan-23 20:03:15  & &  & 0.13$\pm$0.03$^{2}$ & -- & 47.741 & 47.654 & 1.2\\
(90482) Orcus  & 2007-May-30 04:35:16      & &  & 0.36$\pm$0.02$^{1,2}$ &29.27$\pm$2.25$^{1,2}$ & 47.779 & 47.539 & 1.2 \\
(120347) Salacia & 2006-Dec-03 14:34:48    & &  & 0.55$\pm$0.02$^{4}$ & 36.60$\pm$3.70$^{4}$ & 43.819 & 43.390 & 1.2\\
(136108) Haumea & 2007-July-13 11:01:22    & & & no data & 15.83$\pm$1.20$^{2}$ & 51.152 & 50.926 & 1.1\\
\hline
\end{tabular}
\end{center}
}
\tablebib{(1) Stansberry et al. 2008; (2) Mueller et al., 2012;  (3) Brucker et al. 2009; (4) Stansberry et al. 2012. }
\end{table*}

\subsection{Spitzer-MIPS observations}

About 75 TNOs and Centaurs in the  {\it TNOs are Cool} programme were also
observed by the Spitzer Space Telescope (Werner et al. 2004) using the
Multiband Imaging Photometer for Spitzer (MIPS; Rieke et al. 2004). Forty-three
targets were detected at a useful signal-to-noise (S/N $>$ 3) in both the 24 $\mu$m
and 70 $\mu$m bands of that instrument. As was done for our Herschel
programme, many of the Spitzer observations utilised multiple AORs for a
single target, with the visits timed to allow subtraction of background
confusion. The absolute calibration, photometric methods, and colour corrections for the MIPS data are described in Gordon et al. (2007),
Engelbracht et al. (2007) and Stansberry et al. (2007). Nominal
calibration uncertainties are 2\% and 4\% in the 24 and 70 $\mu$m
bands, respectively. To allow for additional uncertainties that
may be caused by the sky-subtraction process, application of colour
corrections, and the faintness of TNOs relative to the MIPS stellar
calibrators, we adopted uncertainties of 3\% and 6\% as done
previously for MIPS TNO data (e.g. Stansberry et al. 2008). The
effective monochromatic wavelengths of the two MIPS bands are 23.68
and 71.42 $\mu$m data. The 24 $\mu$m data band, when combined with the MIPS
or PACS 70 $\mu$m data, provides strong constraints on the colour temperature of the target spectrum, and for the temperature distribution across the surface. In general, Herschel data
alone provide only weak constraints on this temperature distribution,
so the MIPS 24 $\mu$m data are particularly valuable in this respect.

Spitzer flux densities for the nine targets are given in Table~\ref{fluxSpitzer}.
The new and re-analysed flux densities are based on a new reduction of the data using
updated ephemeris positions for the targets (Mueller et al. 2012).
The updated positions sometimes differ by 10\arcsec\ or more
from those assumed by the time of the observation. The ephemeris information is also
used in reducing the raw 70 $\mu$m data, for generating the sky background
images and for accuratly placing of photometric apertures (especially
important for the Classical TNOs which are among the faintest objects
observed by Spitzer).

\subsection{WISE observations for 10199 Chariklo}

For the Centaur 10199 Chariklo we combined our Herschel observations with the Spitzer and WISE ones.
The Widefield Infrared Survey Explorer (WISE, Wright et al. 2010) was launched in December of 2009, and it spent over a year imaging the entire sky in four bands, W1, W2, W3, and W4, centred at 3.4, 4.6, 12, and 22 $\mu$m, respectively. The WISE Preliminary Release includes data from the first 105 days of WISE survey observations - that is, from 14 January 2010 to 29 April 2010 - that were processed with initial calibrations and reduction algorithms. We get from the WISE catalogue the observations of the Centaurs 10199 Chariklo, which was observed between 16 February 2010 UT=06:08:49  and 17 February 2010 UT=09:08:30.\\
We extracted the WISE observed W3 and W4 magnitudes and converted them via the Vega-spectrum into fluxes. Due to the red colour of Chariklo (compared to the blue calibration stars), there is an additional correction needed (see Wright et al.\ 2010) and the W3-flux has to be increased by 17\%, and the W4-flux has to be lowered by 9\%.
It is also required to apply a colour correction, which we calculated via a TPM prediction of the spectral energy distribution (SED) of Chariklo (corresponding roughly to a black body temperature of slightly above 100\,K). The correction factors are 2.72 ($\pm$ 15\%) in W3 and 1.01 ($\pm$ 1\%) in W4. The large error for the W3 colour correction is due to the uncertain shape of the SED at these short wavelengths. We also added a 10\% error for the absolute flux calibration in  the W3 and W4 bands, an error that was estimated from the discrepancy between some red and blue calibrators (Wright et al.\ 2010), and we combined all errors quadratically. The final mono-chromatic flux densities at the WISE reference wavelengths 11.56 (W3) and 22.09\,$\mu$m (W4) are reported in Table~\ref{fluxSpitzer}.


\begin{table*}
\caption{Radiometric diameters, geometric albedos, and beaming factor or thermal inertia ($\Gamma$) for the objects sample.}
\label{results}
\begin{center}
\begin{tabular}{l|l|l|l} \hline
\hline
Objects & Diam. (km) & albedo (\%) & $\eta$ or $\Gamma$ ($J m^{-2} s^{-0.5} K^{-1}$) \\ \hline \hline
(2060) Chiron (all data)           &   215.6$\pm$9.9  &  16.7$^{+3.7}_{-3.0}$   & 0.95$^{+0.09}_{-0.10}$\\
(2060) Chiron (MIPS \& PACS) & 210.0$^{+11.0}_{-9.6}$ & 17.6$\pm$0.4   & 0.91$^{+0.09}_{-0.10}$  \\
{\bf (2060) Chiron (TPM)} & \bf{218$\pm$20} & \bf{16.0$\pm$3.0 } & \bf{$\Gamma <$ 3 (smooth); 5-10 (rough)}\\ \hline
(10199) Chariklo(MIPS \& Herschel) &     236.8$\pm$6.8 &   3.7$^{+1.0}_{-0.8}$ & 1.17$^{+0.05}_{-0.08}$ \\
(10199) Chariklo( MIPS \& PACS) &     240.8$\pm$7.2 &   3.5$\pm$0.8 & 1.20$^{+0.06}_{-0.08}$ \\
(10199) Chariklo(WISE \& Herschel)& 231.3$\pm$7.3 &  4.0$^{+1.0}_{-0.8}$& 1.07$^{+0.12}_{-0.07}$  \\
(10199) Chariklo(WISE \& PACS)& 235.1$\pm$6.1  &  3.9$^{+0.9}_{-0.7}$ & 1.12$\pm$0.08 \\
{\bf(10199) Chariklo (TPM)} & \bf{248$\pm$18} & \bf{3.5$\pm$1.0}  & \bf{ $\Gamma$ = 3--30} \\
\hline
{\bf(38628) Huya (all data) }             & \bf{458.0$\pm$9.2} & \bf{8.3$\pm$0.4} &   \bf{0.93$\pm$0.02 }\\
(38628) Huya (MIPS \& PACS) & 458.8$\pm$9.2  & 8.3$\pm$0.4 &   0.93$\pm$0.02 \\ \hline
(50000) Quaoar  (all data)       &  1036.4$\pm$30.8  & 13.7$^{+1.1}_{-1.3}$  & 1.66$\pm$0.08 \\
{\bf (50000) Quaoar  (MIPS \& PACS)}   & \bf{1073.6$\pm$37.9}  & \bf{12.7$^{+1.0}_{-0.9}$}  & \bf{1.73$\pm$0.08} \\ 
(50000) Quaoar (TPM) & 1082$\pm$67 & 12.0$\pm$2.0 & $\Gamma$= 2--10  \\ \hline
(55637) 2002 UX25 (all data)  & 692.0$\pm$23.0 & 10.7$\pm$0.8   &   1.07$\pm$0.05 \\
{\bf (55637) 2002 UX25 (MIPS \& PACS)}  & \bf{697.2$^{+23.0}_{-24.5}$ } & \bf{10.7$^{+0.5}_{-0.8}$}   &   \bf{1.07$^{+0.08}_{-0.05}$ } \\ \hline
{\bf(84522) 2002 TC302 (MIPS \& PACS)}        & \bf{584.1$^{+105.6}_{-88.0}$} &  \bf{11.5$^{+4.7}_{-3.3}$} & \bf{1.09$^{+0.37}_{-0.25}$ }\\ \hline
{\bf(90482) Orcus (all data)}           & \bf{958.4$\pm$22.9} &   \bf{23.1$^{+1.8}_{-1.1}$}  & \bf{0.97$^{+0.05}_{-0.02}$} \\
(90482) Orcus (TPM) & 967.6$\pm$63 & 22.9$\pm$3.0 & $\Gamma$ = 0.4-2.0 \\
(90482) Orcus (MIPS \& PACS) & 957.7$\pm$24.1 &   23.7$^{+1.2}_{-1.7}$  & 0.98$^{+0.05}_{-0.04}$ \\ \hline
(120347) Salacia (all data)        & 874.2$\pm$32.0 &    4.7$^{+0.5}_{-0.3}$  & 1.12$\pm$0.05 \\
{\bf (120347) Salacia (MIPS \&PACS)} & \bf{901.0$\pm$45}  &   \bf{4.4$\pm$0.4}   & \bf{1.16$\pm$0.03}  \\ \hline
{\bf(136108) Haumea (all data}) & \bf{1239.5$^{+68.7}_{-57.8}$}   &   {\bf 80.4$^{+6.2}_{-9.5}$} &  {\bf 0.95$^{+0.33}_{-0.26}$} \\ \hline
\end{tabular}
\end{center}
\tablefoot{For each target, several solutions are presented, obtained from the NEATM fit of (i) all data,  (ii) all data except SPIRE. For Chariklo more cases including the WISE data are considered (Herschel means PACS and SPIRE data). For Chiron, Chariklo, Orcus, and Quaoar, solutions from the thermophysical model (TPM) are also presented. Preferred results for each target are in bold.}
\end{table*}

\section{Thermal modelling}

\subsection{NEATM model}

We used the hybrid standard thermal model (hybrid STM) to fit the measured Spitzer MIPS, and Herschel PACS, and SPIRE fluxes (Stansberry et al. 2008). 
The STM (cf. Lebofsky et al. 1986, Lebofsky \& Spencer. 1989, and references therein)
assumes a smooth, spherical asteroid, which is not rotating and/or has zero thermal inertia, and which is observed at zero phase angle. The sub-solar temperature T$_{SS}$ is
\begin{equation}
 T_{SS}= \left(\frac{(1-A) \times S_{\odot}}{\epsilon \eta \sigma r^2} \right)^\frac{1}{4}
\end{equation}
where A is the Bond albedo, S$_{\odot}$ is the solar constant, $\eta$ the
beaming factor, $\sigma$ the Stefan-Boltzmann constant, r the
heliocentric distance, and $\epsilon$ the emissivity.
The beaming factor $\eta$ adjusts the sub-solar temperature and is fixed to 0.756 from calibrations using
the largest main belt asteroids.

The hybrid STM  used in this paper is equivalent to the near-Earth asteroid thermal model (NEATM, Harris 1998), except that it assumes zero phase angle, which is a good approximation for TNOs. The difference to the STM is that in the NEATM $\eta$ is fitted with the data instead of using a single canonical value.
NEATM was originally developed for near Earth objects (NEOs) but is applicable to all atmosphereless bodies and is being used in companion papers by Mommert et al. (2012), Vilenius et al. (2012), and Santos-Sanz et al. (2012).\\
The NEATM model assumes a spherical shape of the body. The
temperature distribution follows instantaneous equilibrium of a
smooth surface with solar input. The $\eta$ parameter represents empirically 
the combined effects of thermal inertia and surface roughness.  A grey emissivity ($\epsilon$ = 0.9), constant for all wavelengths, was assumed when calculating the local temperatures and monochromatic fluxes.
The model also requires
the choice on a phase integral q, which we assumed to be
related to the geometric albedo p$_V$ through q = 0.336*p$_{V}$ + 0.479 (Brucker et al. 2009). Diameter and geometric albedo are further related through the relationship:
\begin{equation}
 D= \frac{2 \times 10^{V_{\odot}/5} \times 10^{-H_V/5}}{\sqrt{p_V}} \times 1 AU/km
\end{equation}
where D is the asteroid diameter in km, p$_V$ the geometric albedo, H$_V$ the absolute magnitude (in the V band), and V$_{\odot}$ = -26.76$\pm$0.02 is the Sun V--band magnitude (Bessel et al. 1998). 
\\
For the TNOs, we used the V-band absolute magnitude (H$_V$) values resulting from the average of all the H$_V$ values published in the literature (see Table~\ref{fluxHerschel}). For the centaurs Chiron and Chariklo, considering that their H$_V$ magnitude varies over time, we used the H$_V$ estimation obtained closest to the Herschel thermal observations (see the Sections 4.1 and 5 for more details).\\
To get the solution of the three parameters D, p$_V$, and $\eta$ we considered the minimum $\chi^2$ value between the hybrid standard thermal model and the observed fluxes,  accounting for their individual 1$-\sigma$
error bars. When only upper limits were available, they
were treated as non-detections (i.e. zero flux) with a 1-$\sigma$ uncertainty
equal to 1-$\sigma$ the upper limit.  Uncertainties on the fitted parameters
were obtained using a Monte Carlo approach (see Mueller et al. 2011) in which 1000
synthetic datasets were randomly generated using the uncertainties
in the measured fluxes and in the H magnitude.
The diameter, albedo, and $\eta$ of each of the 1000 synthetic objects were
fitted, and their distributions were used to determine the error bars
on these parameters. As outlined in Mueller et al. (2011), these
distributions are generally not Gaussian (especially the albedo
distribution). Therefore, we adopted the median of the Monte-
Carlo results as the nominal value, and asymmetric error bars to
include the 68.2 \% of the results, as these authors did.

\subsection{Thermophysical model}

We also used a thermophysical model (TPM, Lagerros 1996, 1997, 1998; M\"uller \& Lagerros 1998), where the temperature distribution is calculated for the given illumination and observing geometry, rotation axis, and period.
If the spin-vector properties were not available we included calculations for rotation periods of 6 and 24\,h and various spin-vector orientations: pole-on, equator-on (obliquity 0 and 180 deg), and rotation axes pointing to the ecliptic north and south poles. The TPM assumes a surface roughness (M\"uller \& Lagerros 2002, Lagerros 1998), while the eta-driven models work with smooth surfaces.\\
Here we considered different levels of roughness ranging from r.m.s. values for the surface slopes from 0.1 (relatively smooth) to 1.0 (very rough surface) in the TPM analysis.
Another difference to the simpler models: we used different wavelength-dependent emissivity models (see M\"uller \& Lagerros, 1998, 2002): (i) emissivity of 0.9, constant over all wavelengths; (ii) the default emissivity model with values close to 0.8 in the submm/mm wavelength range (based on Ceres/Pallas observations); (iii) the Vesta-like emissivity model with values close to 0.6 in the submm/mm wavelength range (based on Vesta observations, M\"uller \& Lagerros, 2002). See Section 13 for a detailed discussion about these wavelength--dependent emissivity models and about the emissivity effects on Centaurs and TNOs. \\
In Table~\ref{results} we report the results from the NEATM and TPM modelling. For each target, the NEATM model was run on all the data available (so MIPS, PACS, and SPIRE fluxes, and, for Chariklo, also the WISE data), and also on the MIPS and PACS fluxes alone. In this last solution we exclude the  longer wavelengths fluxes from SPIRE, wavelengths that are in  the Rayleigh-Jeans regime of the objects SED and that are most affected by emissivity effects. As can be seen from Table~\ref{results}, the NEATM results from the fit of all the data and all the data except SPIRE fluxes in general give similar results within the uncertainty values.


\section{Centaur (2060) Chiron}

\begin{figure}
   \centering
\includegraphics[width=9.4cm,angle=0]{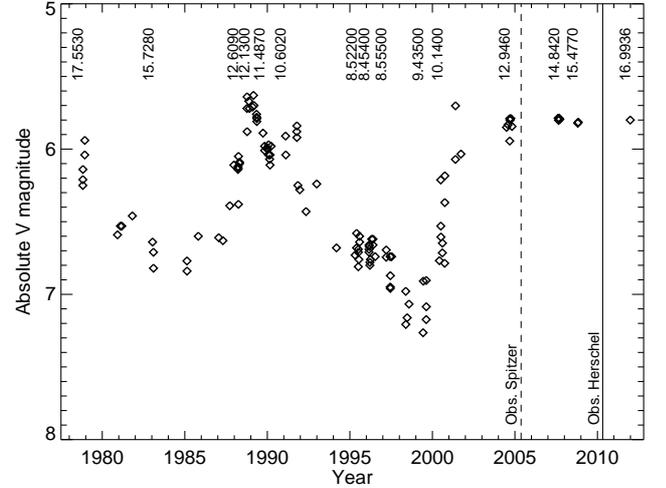}
      \caption{Absolute magnitude of 2060 Chiron versus time (updated version of Fig. 2 from Belskaya et al. 2010). The heliocentric distance (in AU) is indicated on the top. The two vertical lines represent the dates corresponding to the observations obtained with the Spitzer (dashed line) and Herschel telescopes.}
         \label{Chiron_mag}
   \end{figure}

\begin{figure}
   \centering
\includegraphics[width=9.4cm,angle=0]{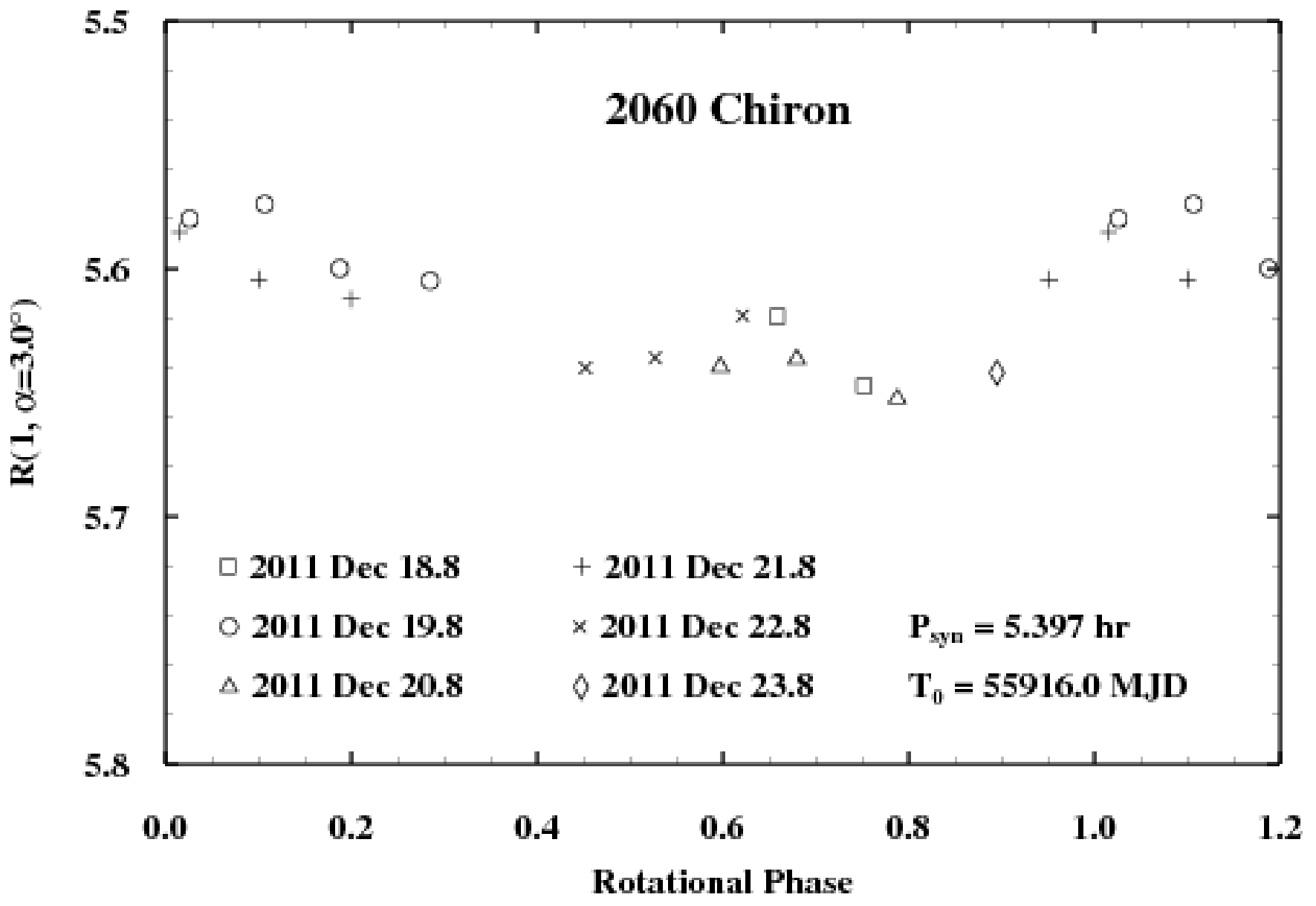}
      \caption{Lightcurve of 2060 Chiron from the 18--24 December 2011 observations at the Calar Alto 1.2 m telescope.}
         \label{Chiron_LC}
   \end{figure}
\begin{figure}
   \centering
\includegraphics[width=9.4cm,angle=0]{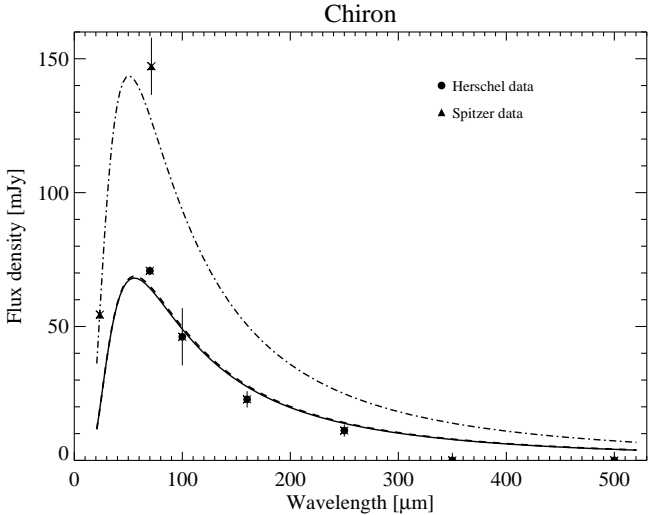}
      \caption{NEATM thermal model of Chiron: model of the MIPS, PACS, and SPIRE data scaled to the heliocentric distance corresponding to the Herschel (solid line) and to the Spitzer (dash--dotted line) observations; dashed line: model of the MIPS and PACS data scaled to the heliocentric distance corresponding to the Herschel observations.}
         \label{chiron}
   \end{figure}

\subsection{Chiron's images in the visible range: H$_V$ value and coma search}

For the centaur 2060 Chiron, the choice of its H$_V$ magnitude is delicate, since this object is subject to cometary activity. It was indeed alternatively labelled as comet
95P/Chiron. It displays considerable variation in its total (nucleus + coma) brightness with time. In particular, we do not have photometric observations of this Centaur close in time to the Herschel ones, and even if we had so, there remains the issue of correcting the H$_V$ magnitude for a possible coma contribution. Here
we make use of the study of  Belskaya et al. (2010), who investigated the photometric properties of Chiron versus
time and heliocentric distance (see especially Fig. 2 of their paper).
Putting together all the photometric data available and analysing in particular those taken in  2004-2008, Belskaya et al. (2010) derived a linear phase coefficient of 0.06$\pm$0.01 mag/deg. The absolute magnitude shown
in Fig. 2 of their paper was calculated from the original photometric data (mainly from Bus et al. (2001), Duffard et al. (2002), Bagnulo et al. (2006), plus Belskaya et al. own data) using this phase coefficient value and a V-R colour of 0.36 (from Barucci et al. 2005). These observations, which were acquired before 2008, show that Chiron's absolute V magnitude was almost constant over the
2004--2008 period, with an H$_V$ value (5.82$\pm$0.07) comparable to the one found on 1989, when the object was at an heliocentric distance of 12.5 AU, and it showed a peak of cometary activity (see Fig.~\ref{Chiron_mag}). \\
No more recent results on Chiron photometry were published. Some observations of Chiron since 2008, mostly from amateur observers, are reported in the minor planet database. Derived H$_V$ ranges from 5.73 to 5.98, consistent with the values derived from Belskaya et al. (2010), but, considering that no details are given about the photometric quality of these observations, we prefer not to include these data in our analysis.

To understand if the Chiron's absolute magnitude is still so bright, our team carried out photometric observations of Chiron in the R filter at the 1.2 m telescope in Calar Alto, Spain on 18 to 24 December 2011, when Chiron
was at R$_h$ = 16.9936 AU, and at a phase angle of 3 degrees. The mean value of the seventeen individual observations taken over the seven nights gives an R magnitude R(1,1,$\alpha$=3) =  5.62$\pm$0.03. Using (V-R)=0.36 and the same phase coefficient obtained by Belskaya et al. (2010), we derive an absolute magnitude H$_V$=5.80$\pm$0.04, consistent with the mean value found over 2004-2008 (Fig.~\ref{Chiron_mag}). \\
The data from the Calar Alto 1.2m telescope, folded with the best-fit period, are shown in Fig.~\ref{Chiron_LC}. They indicate a lightcurve
with period P$_{syn}$ = 5.40$\pm$0.03 hour and amplitude equal to 0.06-0.07 magnitude.
The P$_{syn}$ here obtained is shorter and different compared to that of 5.9178 hours published by different authors (i.e Sheppard et al. 2008), and the analysis of the period is strongly limited by the small number of data available during the December 2011 observations.

Chiron's lightcurve amplitude has relevance to the activity level. Luu \& Jewitt (1990) found that it varies
from $\Delta$m = 0.09 magnitude when Chiron does not show any sign of activity down to $\Delta$m = 0.03 magnitude when it is at its brightest. Chiron's lightcurve variations may be explained by the dilution of the lightcurve by an optically thin coma (Luu \& Jewitt, 1990).
In this framework, the December 2011 amplitude of 0.06-0.07 mag appears to be an intermediate
value, and suggests some moderate cometary activity at that epoch. A similar modulation of the water ice
absorption signatures on the surface has been reported (Forster et al. 1999, Luu et al. 2000), with stronger
contrast of the H$_2$O absorption features when the object does not show any sign of activity.

\begin{figure*}
   \centering
\includegraphics[width=17cm,angle=0]{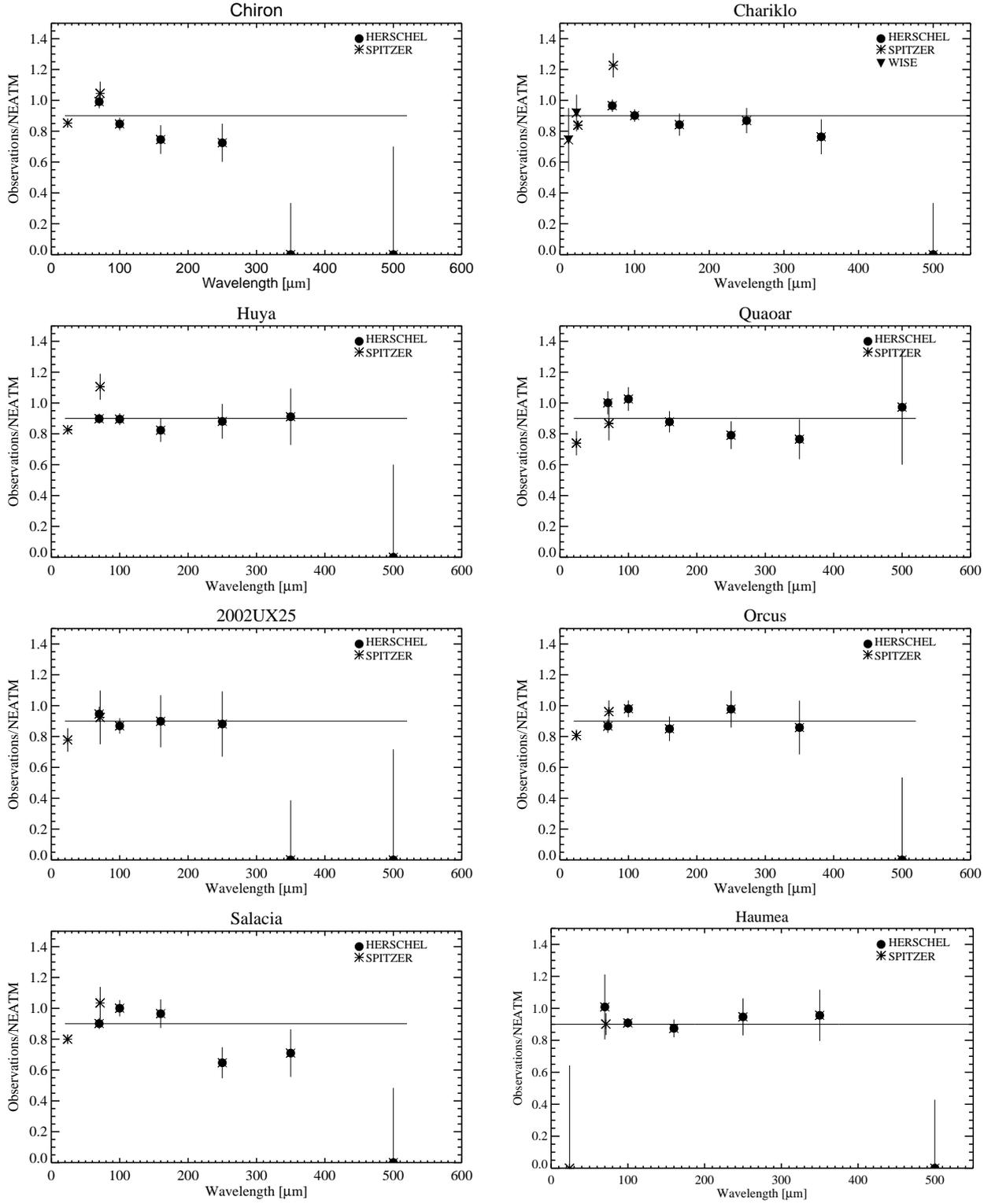}
      \caption{Observed fluxes of the targets divided by the NEATM model that best fit the Herschel and Spitzer data. In the model the emissivity is assumed constant for all wavelengths ($\epsilon$ =0.9).}
         \label{emi_all}
   \end{figure*}

Considering all measurements over 2004-2011, we adopt H$_V$=5.81$\pm$0.08 for the mean total magnitude of Chiron. This indicates a high flux from the Centaur, comparable to that of its activity peak during 1989. \\
Which then is the reason of the Chiron's almost constant high brightness during the latest seven years? One possible answer is that cometary activity is still present on Chiron. To test this hypothesis, we re-analysed some deep Chiron images taken in 2007--2008 (reported by Belskaya et al. 2010) in the Bessel R filter with the FORS1 instrument at the ESO-VLT telescope. The analysis of those images with the $\sum$Af function (Tozzi et al., 2007), which describes the dust albedo (A) multiplied by the total area covered by the solid particles in an annulus of a given radius and unitary thickness, allows us to detect a faint evolved coma (Tozzi et al., 2012).  Its $Af\rho$ resulted to be 650$\pm$110 cm, constant within errors.  Assuming that at the time of the Herschel observation the coma of Chiron had the same Af$\rho$ value, the contribution of the coma into an aperture of 10 arcsec is less than 10\% of the flux of the nucleus. This means that the measured Chiron flux in the visible is almost entirely due to the nucleus, with the coma contributing only to $\sim$ 0.11 mag of the Chiron total magnitude previously determined (5.81$\pm$ 0.08). Correcting for this small coma contribution we find that the nuclear H$_V$ magnitude of Chiron was around 5.92$\pm$0.20, where we adopt a conservative error bar to take into account the non-simultaneousness between visual and thermal observations and the errors in the coma contribution estimation. \\
Nevertheless there is still the possibility that an unresolved evolving  coma contributes to Chiron's
brightness. In the December 2011 observations we performed, however, the coma, if present, was below our detection limit.

\subsection{Chiron: NEATM model}

Assuming $H_v$=5.92 §$\pm$0.20 for Chiron nucleus, the NEATM model of the revised Spitzer-MIPS and Herschel PACS and SPIRE data (Fig.~\ref{chiron}) gives a diameter of 215.6$\pm$9.9 km, and a geometric albedo of 16.7$^{+3.7}_{-3.0}$ \% (Table~\ref{results}). \\
Very similar results are obtained from the TPM modelling with a constant emissivity of 0.9 (diameter of 210.8 km and $p_v$ = 17.3\%). From the analysis of the Spitzer-MIPS data, Stansberry et al. (2008) report a similar value for the diameter (233$\pm$14 km),
but a much lower albedo value (7.5\%), which results from their assumed $H_v$ value (6.58).  However, Figure~\ref{Chiron_mag} indicates that the $H_v$  magnitude at the time of the Spitzer observations must be close to the value estimated during the Herschel observations. \\
 At the other extreme, one might interpret the minimum brightness level in Fig.~\ref{Chiron_mag}
(H$_V$= 7.26), reached in June 1997, as indicative of the true nuclear magnitude. Doing so we would obtain D = 206 km and p$_V$ = 5.3 \%. Altogether we conclude that Chiron's diameter from NEATM modelling, and with constant emissivity value, is in the range 196-225 km (including a 10 km error bar).
The albedo is much more loosely constrained at $\sim$5-17 \%, but we favour the higher albedo given our analysis of the H$_V$ value close to the Herschel observations and of the relative coma contribution. Nevertheless, if the nuclear H$_V$ magnitude is really $\sim$5.92 during the Herschel observations, it remains to be understood what caused the factor of 3.5 increase in the nucleus flux from 1997 to 2004-2011, and one speculation is that it results from surface ``repainting'' following an activity outburst.

\subsection{Chiron: TPM model}

 Chiron shows strong emissivity effects for wavelengths beyond 100 $\mu$m (see Fig.~\ref{emi_all}). We model the data with the TPM model with both constant and wavelength dependent emissivity. We find that the  TPM model with constant emissivity is not working very well, and the long-wavelength data are poorly matched by the model prediction, because of this significant decrease in the emissivity versus wavelengths. We then run a TPM model with wavelength-dependent emissivity derived from the SED of Vesta, which gives lower $\chi^2$ values and so is a better fit. Nevertheless, the TPM model gives a degeneracy between surface roughness and thermal inertia: a smooth surface requires a low value of the thermal inertia ($<$ 3 in SI units ($J m^{-2} s^{-0.5} K^{-1}$)),  while a high roughness would also require higher thermal inertia in the range 5-10 in SI units. \\
 \begin{figure}
   \centering
\includegraphics[width=7.2cm,angle=90]{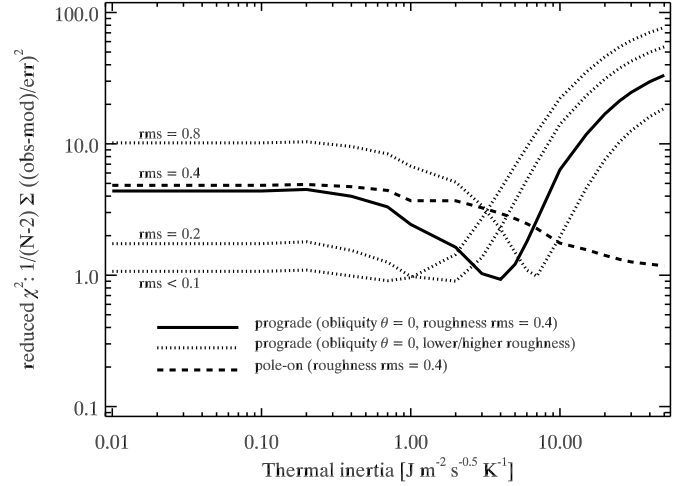}
      \caption{TPM analysis of Chiron thermal fluxes: $\chi^2$ test for various spin-vector orientations. $\theta$=0$^o$ indicates a geometry with the spin-axis perpendicular to the orbit plane and prograde rotation. The lines represents model solutions with different levels of roughness.}
         \label{chiron_tpm}
   \end{figure}
Assuming our best guess of H$_V$, this model predicts 0.13 $<$ p$_V$ $<$ 0.19 and a diameter ranging between 198-238 km for the different spin-vector orientations, different levels of roughness and the aforementioned corresponding thermal inertia values. There are not enough data to distinguish between the different spin-axis orientations, even if the obliquities 0 and 180 degrees both give very low reduced $\chi^2$ minima (see Fig.~\ref{chiron_tpm}). The pole-on case does not constrain the thermal inertia significantly, and the corresponding size and albedo solutions are  D =198-213 km, and p$_V$=0.17--0.19 for the high roughness case, and  D = 224-238 km, p$_V$=0.13-0.15 for the low roughness case.

Finally, considering both NEATM and TPM models with constant and wavelength-dependent emissivity, respectively, Chiron's diameter appears to be relatively well constrained from
the combined Herschel-Spitzer observations. The 198-238 km solution range is consistent with the larger size solutions previously determined from infrared or mm-range radiometry
(Lebofsky et al. 1984, Campins et al. 1994; Fernandez et al., 2002; Sykes \& Walker, 1991; Groussin et al., 2004; Jewitt \& Luu, 1992; Altenhoff \& Stumpff, 1995; Stansberry et al. 2008). In fact, these  measurements carried out from 1983 to 2000 give a Chiron diameter ranging from 126 to 228 km, as summarised by Groussin et al. (2004, see Table 3 of their paper). Our size value is also consistent with occultation results (Buie et al. 1993, Bus et al. 1996), which give a lower limit of 180 km for Chiron's diameter, thus excluding the smaller diameter solutions.

\subsection{Look for activity in the Chiron's PACS images}

Beyond determining Chiron's size, albedo, and thermal
properties, we analysed the Chiron Herschel-PACS images at 70 and
100 $\mu$m to look for a coma pattern. To do this we processed the
Chiron images using the TNOs--are--Cool HIPE modified script for
point-like sources (Santos-Sanz et al. 2012). We extracted the
radial profile of these images by integrating the fluxes in annuli
centred on the photocentre and spanning from 0 to 31 pixels in
radius (with an annulus width of one pixel). The FWHM of Chiron
emission was derived by fitting these radial profiles by a Gaussian
function. Finally, we compared these FWHMs with the values obtained
from a mean radial profile extracted from eight standard stars images.
We analysed scan A and scan B images separately. The results
discussed later in this section come from the mean value between 
these two independent measurements.\\
\begin{table}[h]
\caption{FWHM derived for Chiron, Chariklo, and stars images. A and B indicate the scans A and B. For the stars, the FWHM for the two A and B scans is derived from the mean value of the profile analysis of 8 stars.} 
\label{coma}
\begin{center}
\begin{tabular}{lccccc} \hline\\
Object &        Band &     FWHM$_{A}$ &   S/N$_A$ & FWHM$_{B}$ &   S/N$_B$ \\
         & ($\mu$m)  & (\arcsec)      &           & (\arcsec) & \\ \hline \hline
Chiron  & 70      &  5.76$\pm$0.04   & 24  & 5.74$\pm$0.03 & 28 \\
Chariklo  & 70      &  5.85$\pm$0.03   & 30 &  5.80$\pm$0.03   & 40  \\
Stars   & 70      &  5.81$\pm$0.02   & 102 & 5.71$\pm$0.03   & 106 \\
Chiron  & 100     &  6.48$\pm$0.07  &  13 &  6.66$\pm$0.07  &  15\\
Chariklo  & 100     &  6.99$\pm$0.06  &  23 & 6.83$\pm$0.05  &  23\\
Stars   & 100     &  6.88$\pm$0.04  &  39 &  6.86$\pm$0.06  & 43\\ \hline
\end{tabular}
\end{center}
\end{table}
The FWHM from the standard stars images was taken to be
representative of point-like emission. From this analysis
(Table~\ref{coma}), we conclude that the coma is not detected in
the Chiron images at 70 and 100 $\mu$m. Indeed, a broadening of
the radial profile with respect to that of standard
stars images would then have been observed. \\
To derive constraints on the dust production rate, we estimated the maximum
contribution of the coma to the measured signal. Radial profiles
expected for the combination of nucleus and coma emissions were
computed, assuming that the dust number density follows a 1/$r^2$
profile in the coma, where $r$ is the distance to the body centre.
Conservatively, we searched for the linear combination giving FWHM =
FWHM(Chiron)+5$\times$$\sigma$(Chiron) (e.g., at 70 $\mu$m we used
FWHM(Chiron) = 5.75\arcsec~and $\sigma$(Chiron) = 0.04\arcsec, see
Table~\ref{coma}). The relative contributions of the coma to the
observed signals is at most 4.1\% and 3.4\% (5--$\sigma$ level) 
at 70 and 100 $\mu$m, respectively. This sets the maximum dust
fluxes to 0.062 and 0.041 mJy/pixel at the photocentre of the 70
and 100 $\mu$m images, respectively (with pixel sizes of 1.1'' and
1.4'', respectively).

\subsection{Chiron: Upper limits in the dust production rate}

\begin{table*}[ht]
\caption{Dust parameters and production rates for Chiron derived
from PACS 70 and 100 $\mu$m data. } 
\label{dust}
\begin{center}
\begin{tabular}{cllllll} \hline\\
 $Q_{\rm gas}$ & Outgassing & $a_{\rm max}$ & $v_{\rm d}^a$  &  $\alpha$ & $Q_{\rm dust}^{b,c}$ & $Q_{\rm dust}^{c,d}$\\
 (s$^{-1}$)  &            & ($\mu$m)    &  (m s$^{-1}$) &           & (kg
 s$^{-1}$) & (kg s$^{-1}$) \\
\hline \hline  \noalign{\smallskip}
(1) 1 $\times$ 10$^{28}$          &    isotropic    &  0.14   & 89--99  & --3 & $<$ 10--14 & $<$ 6--15 \\
(2) 5 $\times$ 10$^{27}$          &    SSL $<$ 45$^{\circ}$    &  0.47   & 89--143  & --3 & $<$ 14--17 & $<$ 8--18 \\
(3) 3 $\times$ 10$^{28}$          &    SSL $<$ 45$^{\circ}$    & 2.76   &  89--206 & --3 & $<$ 25--29 & $<$ 17--29 \\
(4) 1 $\times$ 10$^{29}$          &    SSL $<$ 45$^{\circ}$    & 9.20   &  89--240 & --3 & $<$ 23--47 & $<$ 22--45 \\
 \hline
\end{tabular}
\end{center}
\tablefoot{$^a$ Dust velocities in the size range $a_{\rm min}$--$a_{\rm max}$ ($a_{\rm min}$ = 0.1 $\mu$m); $^b$ Upper limit (5--$\sigma$) on the dust production rate from 70-$\mu$m PACS data; $^c$ The lower and higher values correspond to carbon and olivine grains, respectively; $^d$ Upper limit (5--$\sigma$) on the dust production rate from 100$\mu$m PACS data.
}
\end{table*}

In order to derive limits on the dust production rate $Q_{\rm
dust}$, we used the model of dust thermal emission applied to the
PACS data of comet C/2006 W3 (Christensen) (Bockel\'ee-Morvan et
al., 2010). The basic principles of this model are given in Jewitt
\& Luu (1990). Absorption cross-sections calculated with the Mie
theory are used to compute both the temperature of the grains,
solving the equation of radiative equilibrium, and their thermal
emission. Complex refractive indices of amorphous carbon and
olivine (Mg:Fe = 50:50) (Edoh, 1983; Dorschner et al., 1995) were
taken as broadly representative of cometary dust. We considered a
differential dust production $Q_{\rm dust}$($a$) as a function of
grain radius $a$, described by the size index $\alpha$. The
size-dependent grain velocities $v_{\rm d}$($a$), as well as the
maximum grain radius $a_{max}$, were computed following Crifo \&
Rodionov (1997). We assumed a nucleus and dust density of 500 kg
m$^{-3}$. Both the maximum grain size and dust velocities
critically depend on the gas production rate at the surface, which
is poorly constrained. \\
An upper limit of 5 $\times$ 10$^{27}$
s$^{-1}$ for the CO production rate was measured in 1998/1999,
when Chiron was at its minimum brightness (Bockel\'ee-Morvan et al., 
2001), whereas a marginal detection (Womack \& Stern, 1999) and
upper limits (Rauer et al.,  1997), all in the range (1--3)
$\times$ 10$^{28}$ s$^{-1}$, were obtained from observations in
1995, just before its perihelion passage on February 1996.  \\
An evidence of a gaseous coma was also obtained from the detection of 
the CN band in the visible, as reported by Bus et al.  (1991), possibly during
an outburst. The outgassing geometry is also uncertain. For the
distant comets 29P/Schwassmann-Wachmann 1 and C/1995 O1
(Hale-Bopp), CO outgassing was observed to mainly be in the Sun's direction (Gunnarsson et al., 2003, 2008). \\
We inferred that, owing to the large size of Chiron, the gas
production rate $Q_{\rm gas}$ should exceed $\sim$ 10$^{28}$
s$^{-1}$ to lift-off dust particles with size $a$ $>$ 0.14 $\mu$m,
assuming that the outgassing is uniform over the surface. For an
outgassing restricted to subsolar latitudes SSL $<$ 45$^{\circ}$,
the maximum size is $a_{\rm max}$ $=$ 0.14 $\mu$m for $Q_{\rm
gas}$ = 1.5 $\times$ 10$^{27}$ s$^{-1}$. Hence, significant gas
production is needed to explain the presence of a faint coma in
the optical data of 2007--2008 (Tozzi et al., 2012). \\
Based on these considerations, and assuming that cometary activity was present at
the time of the Herschel observations, we investigated four
models: 1) uniform outgassing with $Q_{\rm gas}$ = 1 $\times$
10$^{28}$ s$^{-1}$; 2) gas production within subsolar latitudes
SSL $<$ 45$^{\circ}$ with $Q_{\rm gas}$ = 5 $\times$ 10$^{27}$
s$^{-1}$; 3) same as 2), with $Q_{\rm gas}$ = 3 $\times$ 10$^{28}$
s$^{-1}$; 4) same as 2), with $Q_{\rm gas}$ = 1 $\times$ 10$^{29}$
s$^{-1}$. The latter model allows us to investigate possible
higher activity in 2010 with respect to 1995--1999. The values of
$a_{\rm max}$ and the dust velocities for the minimum (0.1 $\mu$m)
and maximum size grains for the four models are given in
Table~\ref{dust}. All results correspond to a CO$_2$ dominated
coma. \\
Similar numbers are obtained considering instead CO
outgassing. The larger grains have sizes from 0.14 to 9.2 $\mu$m
and a velocity of $\sim$ 100 m s$^{-1}$, whereas 0.1--$\mu$m
grains display a range of velocities depending on the gaseous
activity (Table~\ref{dust}). Table~\ref{dust} shows the upper
limits on $Q_{\rm dust}$ derived from the 70 and 100 $\mu$m PACS
data for a size index $\alpha$ = --3.  Values range from 6 to 45
kg s$^{-1}$, depending on the outgassing model and assumed dust
composition. The values are not sensitive to the size index
(within $\sim$$\pm$10\% considering $\alpha$ from --2.5 to --3.5).

These upper limits can be compared to the dust production rate in
2007--2008 that can be derived from the $Af\rho$ value of 650 cm
(Tozzi et al., 2012). Assuming single-size grains, $Af\rho$ is
related to $Q_{\rm dust}$ through (Jorda 1995) :

\begin{equation}
Q_{\rm dust} =  \frac{2}{3} Af\rho \times \frac{\rho_{\rm d} a
v_{\rm d}}{A_{\rm p}}
\end{equation}

\noindent where $a$ is the grain radius, $\rho_{\rm d}$ is the
dust density, and $A_{\rm p}$ is the geometric albedo of the dust.
The derived $Q_{\rm dust}$ is $\sim$ 5 kg s$^{-1}$ for $a$ = 1
$\mu$m, $v_{\rm d}$ = 100 m s$^{-1}$, $\rho_{\rm d}$ = 500 kg
m$^{-3}$, and $A_{\rm p}$ = 0.04. Values from 0.6 kg s$^{-1}$
(model (1)) to 17 kg s$^{-1}$ (model (4)) are derived when
considering the dust size distributions and velocities utilised to
interpret the PACS data (Table~\ref{dust}). Therefore, under the
assumption of a dust albedo of 0.04, a minimum particle size
$a_{\rm min}$ = 0.1 $\mu$m, and a size index $\alpha$ = --3, the
upper limits derived from the PACS 2010-data are within a factor
1.3--10 higher than the dust production rate in 2007--2008 derived
from optical data. We note, however, that the geometric albedo of
the grains might be one order of magnitude lower if the grains are
highly porous (Lacerda \& Jewitt 2012, and references therein). In
this case, the dust production rate in 2007--2008 and the upper
limit for 2010 are comparable. 

Combining our assumptions on the
gas production rate and upper limits on $Q_{\rm dust}$ from PACS
data, the dust-to-gas ratio for Chiron is $<$ 0.04 (conservative
value), i.e., at least two orders of magnitude lower than values
estimated for kilometer-sized cometary nuclei.


\section{Centaur (10199) Chariklo}

\begin{figure}
   \centering
\includegraphics[width=9.4cm,angle=0]{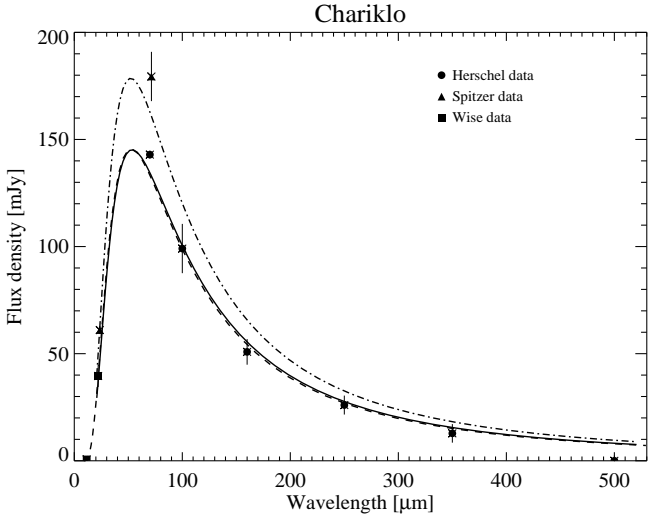}
      \caption{NEATM thermal model of Chariklo: model of the MIPS, PACS, and SPIRE data scaled to the heliocentric distance corresponding to the Herschel (solid line)  and to the Spitzer (dash--dotted line) observations; dashed line: model of the WISE, PACS, and SPIRE data scaled to the heliocentric distance corresponding to the Herschel observations.}
         \label{chariklo}
   \end{figure}

Centaur 10199 Chariklo (1997 CU26) was discovered on February 15,
1997, seven year before its passage at perihelion. As for Chiron,
a strong variation of the absolute magnitude H$_V$ (0.8 mag) over
time was seen in the relatively
small heliocentric distance range covered by Chariklo since its discovery (from 13.8 AU  to the perihelion distance of 13.1 AU),
as summarised by Belskaya et al. 2010 (see Fig. 5 of their paper). Belskaya et al. (2010) reanalysed the
data published by McBride et al. (1999), Jewitt \& Luu (2001), Tegler and Romanishin (1998),
Peixinho et al 2001, Bauer et al (2003), Dotto et al. (2003), Guilbert et al. (2009), and their own data,
finding a linear phase coefficient of 0.06$\pm$0.01 mag/$^o$ in the phase angle range 1.2-4.1$^o$,
that was applied to all these observations to determine the corresponding H$_V$ value. \\
While for Chiron the H$_V$ magnitude becomes brighter after its perihelion passage, Chariklo on the other
becomes fainter after its perihelion passage in 2004. Both Guilbert et al. (2009)  and Belskaya et al. (2010)
report an H$_V$ value of 7.2-7.4, well below the pre-perihelion values (6.46-7.0). We do not have photometric
observations simultaneous to the Herschel ones, so we take an H$_V$ value of 7.4$\pm$0.25 corresponding to the
absolute magnitude estimation closest to our observations (March-June 2008, from Belskaya et al. 2010), and where a conservative large error bar of 0.25 mag is assumed to take possible undetected coma contribution effects and small H$_V$ time variations into account, than cannot be excluded. \\
The decrease in Chariklo absolute magnitude after its perihelion passage may have 2 possible explanations: 1) in the past Chariklo experienced cometary activity that stopped or decreased after its perihelion passage, and as consequence, its absolute brightness decreased; 2) spin axis orientation effects, with possibly pole-on geometry in the 1999-2000 period, and an equatorial view in the 2007-2010 period.

\subsection{Chariklo: NEATM and TPM thermal modelling}

With this H$_V$ value, the diameter derived from the NEATM model of the revised Spitzer-MIPS and Herschel PACS and SPIRE data (Fig.~\ref{chariklo}) is D$_{eff}$= 236.8$\pm$6.8 km, and the geometric visual albedo of 3.5$^{+1.0}_{-0.7}$ \% (Table~\ref{results}). Stansberry et al. (2008), from the analysis of the Spitzer-MIPS data alone, reported a slightly higher diameter (257$\pm$13 km), and a higher albedo (5.8\%), which result from their assumed $H_v$ value (6.66, but according to Fig. 5 from Belskaya et al. (2010) the H$_V$ value close to the Spitzer observations was $\sim$ 7.0). Other Chariklo size determinations come from millimeter wavelength observations (D$_{eff}$=273$\pm$19 km from Altenhoff et al (2001)), or from infrared data (D$_{eff}$=302$\pm$30 km from Jewitt \& Kalas (1998); 236$\pm$12 km from Groussin et al. 2004). 

\begin{figure}
   \centering
\includegraphics[width=6.8cm,angle=90]{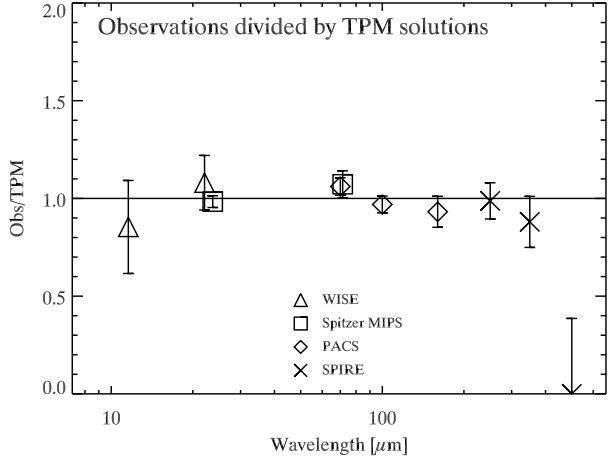}
      \caption{Observed fluxes of Chariklo divided by the TPM model that best fit the data.}
         \label{chariklo_tpm}
   \end{figure}

The Spitzer and Herschel fluxes divided by the NEATM model that best fit the data clearly show a moderate drop-off of the emissivity beyond $\sim$ 300 $\mu$m for Chariklo (Fig.~\ref{emi_all}). To take this effect into account, we ran a TPM model with emissivity decreasing versus wavelength in a Vesta-like manner (M\"uller \& Lagerros, 2002). We also  included in this analysis the fluxes coming from WISE W3 and W4 band observations (Fig.~\ref{chariklo_tpm}). Unfortunatly, the Chariklo rotational period is not known and it is supposed to be long (around 15 or 34 hours according to  Peixinho et al., 2001). In our TPM analysis two options with a rotational period of 6 and 24 hours were run.\\
\begin{figure}
   \centering
\includegraphics[width=6.8cm,angle=90]{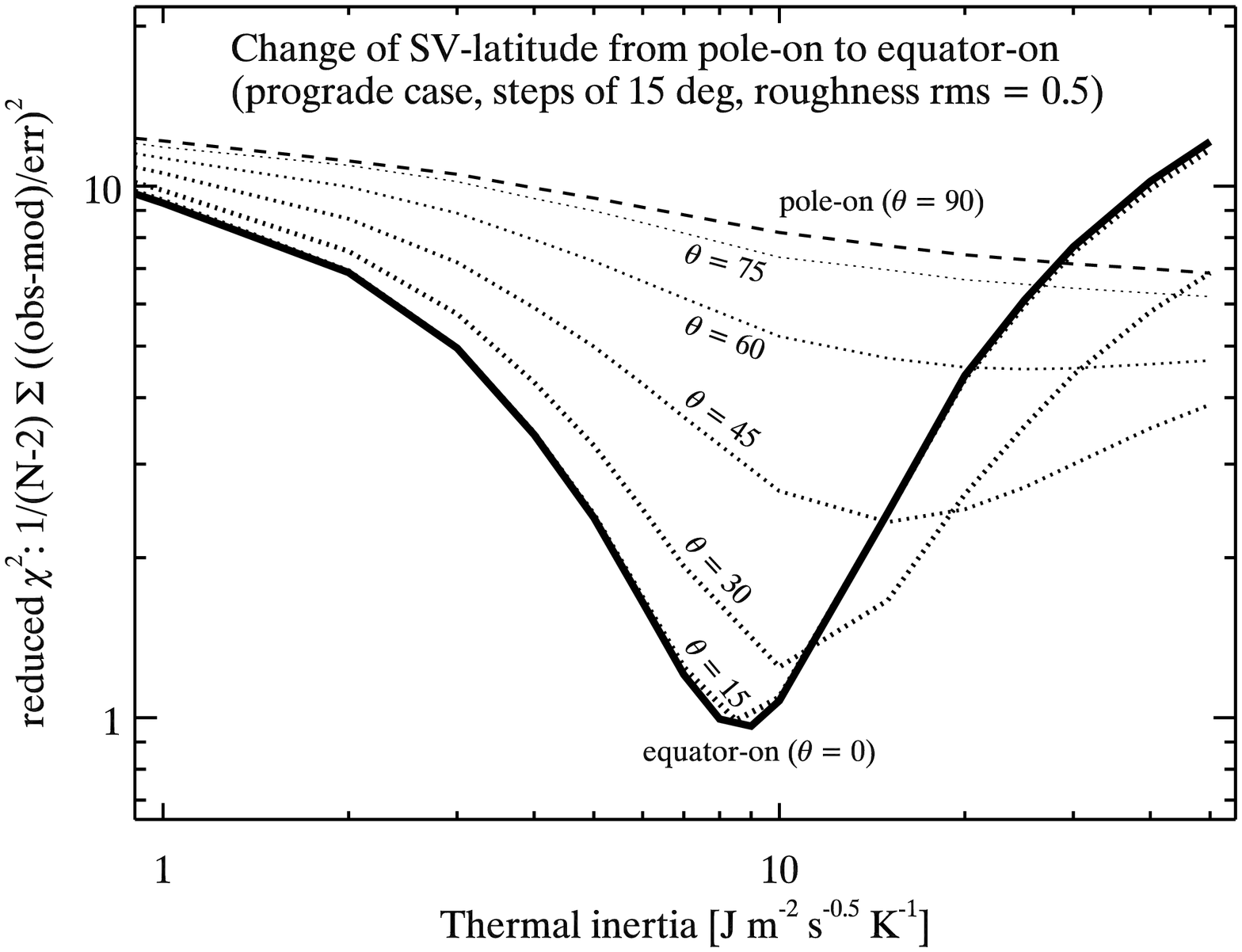}
\includegraphics[width=6.8cm,angle=90]{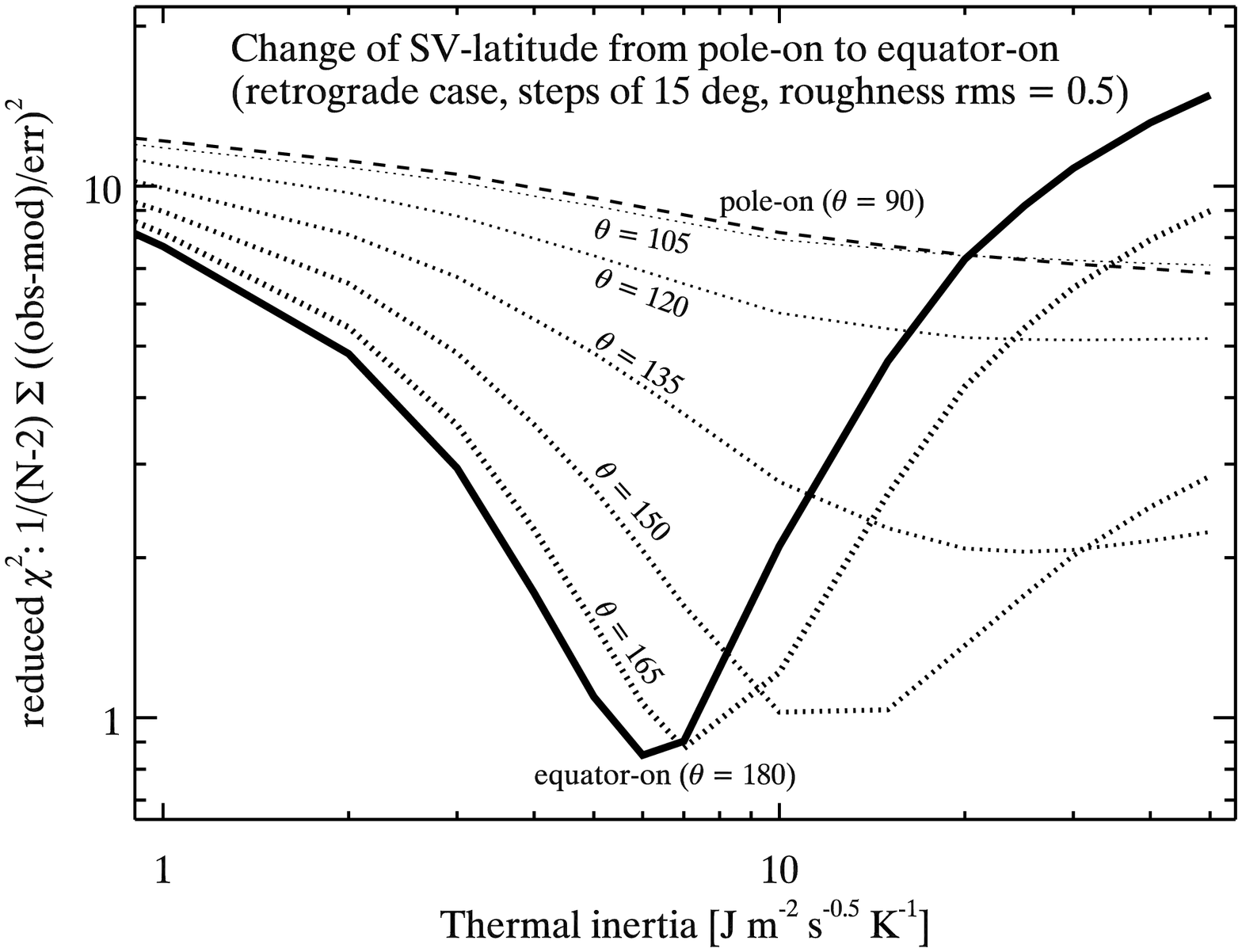}
      \caption{TPM analysis of Chariklo thermal fluxes: $\chi^2$ test for a
wide range of thermal inertias for different spin-vector orientations. Top: moving from a pole-on configuration to equator-on viewing geometry for a prograde rotation ($\theta$ represents the obliquity); Bottom: the same, but for retrograde rotation. Note that the modelling was done for an assumed rotation period of 6 h and intermediate levels of surface roughness.}
 \label{chariklo_chi2}
   \end{figure}
Spherical shape models with different spin-axes were tested to see if the
thermal data can constrain the pole-orientation and how rotation period
and surface roughness would influence the radiometric solution.
Figure~\ref{chariklo_chi2} shows how the match between observed and modelled
fluxes changes (expressed in terms of reduced $\chi^2$ values) when the
spin-axis orientation is modified. The calculations have been done for
a spherical object with an assumed rotation period of 6 h and intermediate
surface roughness levels. The best agreement with all thermal data
simultaneously is found for an equator-on viewing geometry and a thermal
inertia around 9 in case of a prograde rotation or 6 in case of a retrograde rotation. 
Within the acceptable $\chi^2$ values the spin-vector latitude can be given 
with $\beta_{ecl}^{sv} = +66.6^{\circ} \pm 30^{\circ}$ (or $-66.6^{\circ}$ in case of retrograde rotation), close to the obliquity 0$^{\circ}$ or 180$^{\circ}$ orientation for Chariklo which has an orbit
inclination of 23.4$^{\circ}$.
\begin{figure}
   \centering
\includegraphics[width=7.2cm,angle=90]{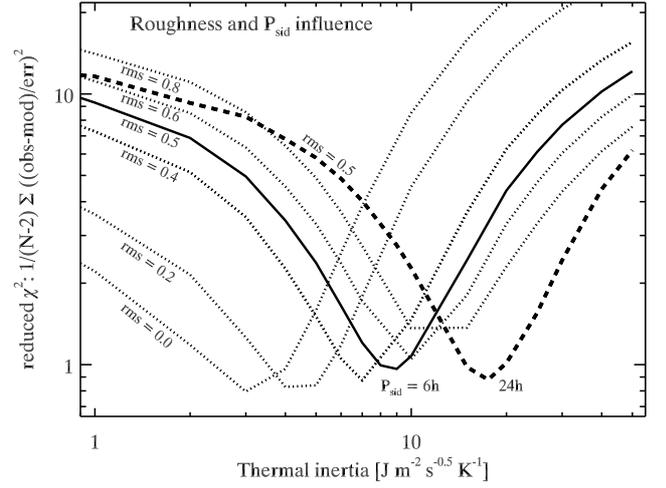}
      \caption{TPM analysis of Chariklo thermal fluxes: $\chi^2$ test for different levels of surface roughness and for two different rotation periods (solid line: 6h, dashed line: 24 h). Note that we assumed here a spherical shape in an equator-on viewing geometry and a prograde rotation.}
         \label{chariklo_tpm3}
   \end{figure}

It is not so easy to establish the possible thermal inertia range. Here the
true rotation period and also the surface roughness plays a role. Slower
rotations would require higher thermal inertias to explain our measurments,
faster rotations lower values. A rougher surface would also lead to larger
values for the thermal inertia, while a smooth surface would require
lower values. This is shown in Fig.~\ref{chariklo_tpm3}.  The TPM model
with the same spin-axis orientation, the same surface roughness, but
a longer rotation period of 24 h (instead of 6 h) would require much higher thermal inertias
of up to 20 (dashed line) to explain the thermal fluxes. The influence of
roughness is shown in the dotted lines: the level of roughness increases
from left to right by steps of 0.2 from a perfectly smooth surface (r.m.s. of
surface slopes 0.0) to a very high level of roughness (r.m.s. of surface
slopes 1.0). \\
In the TPM there is a strong degeneracy between thermal inertia
and roughness: if the object has a high thermal inertia (10 $<$ $\Gamma$
$<$ 30
in SI units) then this would require at the same time a high roughness
(rms-slope $>$ 0.6); if the object has a low thermal inertia (3 $<$
$\Gamma$ $<$ 10
in SI units) then this would require a relatively smooth surface
(rms-slope $<$ 0.5). Our dataset does not allow to break this degeneracy
between
surface roughness and thermal inertia (although the intermediate low
roughness levels are favoured) and it is also not possible to extract
any information about Chariklo's rotation period.\\
The full possible range for the thermal inertia value is 3-30 in SI
units, with the optimum value just below 10 in SI units for a 6 h rotation period or
just below 20 in SI units for a 24 h rotation period.

Finally, the TPM that best fit the data gives 230 $<$ D$_{eff}$ $<$ 266
km and 0.025 $<$ p$_V$ $<$ 0.045 with a 3$\sigma$ confidence level for the
aforementioned range of thermal inertia values. The specified albedo range includes
already the large uncertainties in the H$_V$ magnitude. So Chariklo is definitely the largest Centaur known so far and it is indeed very interesting because of a relatively high thermal inertia value associated to a very low albedo surface. Chariklo surface may be similar to another peculiar object, the Jupiter Trojan 1173 Anchises, which has been found to have a low albedo (2.7\%) and a comparable high thermal inertia (Horner et al., 2012)  for its moderately large heliocentric distance (5 AU).\\
The fact that the TPM model excludes the pole-on solution at the time of the Herschel observations reinforces the assumption made by Belskaya et al. (2010), who speculate that the 2007-2008 observations after the Chariklo passage at perihelion (corresponding to a fainter H$_V$ value) were made at equatorial view, while the 1999-2000 observations, showing a higher H$_V$ value and short term brightness variations, corresponded to a pole-on geometry. Interestingly, Chariklo spectra indicate surface composition heterogeneities, with the presence of water ice absorption bands in 1997-2001 observations (with band depth up to 20\%), bands that where not detected in the 2007-2008 observing campaigns (Brown et al. 1998, Brown \& Koresko 1998; Dotto et al. 2003, Guilbert et al. 2009). If the assumption of pole-on aspect during 1999-2000 observations is correct, then the spectral heterogeneity may be explained as the Chariklo polar surface having larger amounts of water ice compared to the equatorial surface seen later than 2007 (Belskaya et al. 2010).  \\

\subsection{Chariklo: Upper limits in the dust production rate}

We also analysed the Chariklo images obtained with PACS in 2010 as
done for Chiron, to look for the possible presence of a faint
coma. The PSF profile of Chariklo does not reveal the presence of
any coma at 5--$\sigma$ level in the far-infrared range (see
Table~\ref{coma}). From the same method as used for Chiron (Sect.
4.1.1), we determined the maximum dust fluxes (5--$\sigma$) in the
central pixels of the 70 and 100 $\mu$m images to be 0.17 and 0.18
mJy/pixel, respectively. Upper limits on the dust production rate
are higher than those derived for Chiron (Table~\ref{dust}), under
the same assumptions. For example, for olivine grains using the
70--$\mu$m data, which provide more stringent constraints, we find
$Q_{\rm dust}$ upper limits of 28 and 89 kg s$^{-1}$, for models
(1) and (4), respectively. We also re-analysed some Chariklo
images taken in the Bessel R filter with the FORS1 instrument at
the ESO-VLT telescope during 2007-2008 (reported by Belskaya et
al. 2010) to look for cometary activity. The analysis indicates
that no coma was present during the 2007--2008 observations, and
the corresponding 3 $\sigma$ upper limit of the $Af\rho$ value is 
200 cm (Tozzi et al., 2012). The upper limits for $Q_{\rm
dust}$ derived from $Af\rho$ are 0.16 and 4 kg s$^{-1}$ for
parameters of models (1) and (4), respectively (Table~\ref{dust}),
and more stringent than the values derived from the PACS
observations.
 \\
Considering our results from NEATM model with fixed emissivity and from the TPM model with emissivity varing with lambda in a Vesta like fashion we conclude that the size of Chariklo must be within 230 and 275 km, and that its geometric albedo is very low and comprised between 2.8 and 4.5 \%. Future observations helping in determining the Chariklo rotational period and its pole orientation will be very helpful in refining the size of Chariklo.

\section{Plutino (38628) Huya}

\begin{figure}
   \centering
\includegraphics[width=9.4cm,angle=0]{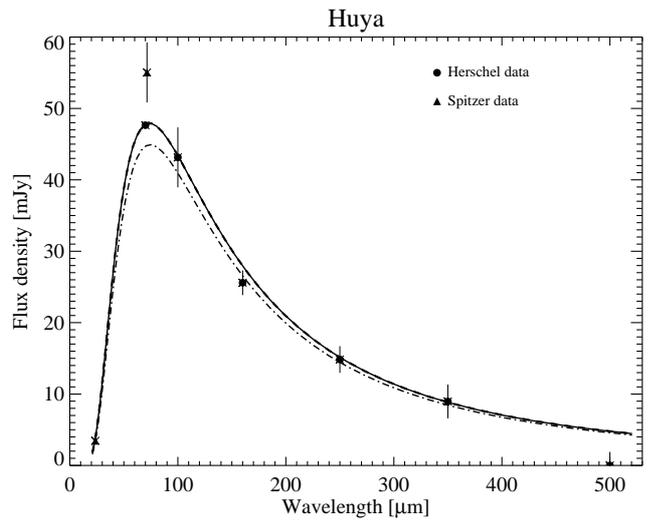}
      \caption{NEATM thermal model of Huya: model of the MIPS, PACS, and SPIRE data scaled to the heliocentric distance corresponding to the Herschel (solid line) and to the Spitzer (dash--dotted line) observations. The model of the MIPS and PACS data (dashed line), scaled to the heliocentric distance corresponding to the Herschel observations, is indistinguishable from the solid line.}
         \label{huya}
   \end{figure}

Huya is a plutino discovered in 2000. Lazzarin et al. (2003) reported the presence of two absorption bands in the visible spectra of Huya and 2000 GN171. These bands are similar to those seen on some primitive main belt asteroids and attributed to aqueous alteration products. Nevertheless, these features have never been confirmed (de Bergh et al. 2004; Fornasier et al. 2004a; Alvarez-Candal et al. 2008). Huya spectrum shows a
generally featureless behaviour in the infrared (Brown et al. 2000, 2007; Licandro et al. 2001), but water-ice bands have been reported by Alvarez-Candal et al. (2007).
These differences in the visible and NIR spectra were interpreted as due to surface heterogeneities (de Bergh et al. 2004, Alvarez-Candal et al., 2007). 

Mommert et al. (2012), from combined Herschel-PACS and Spitzer observations, obtained a diameter of  438.7$\pm$26 km, a p$_V$ = 8.1$\pm$1.1 \%, and a beaming factor of 0.89$\pm$0.06, while Stansberry et al. (2008), on the basis of the Spitzer -MIPS observations alone, derived a larger diameter (D = 533$\pm$25 km) and a lower albedo value (p$_V$ = 5.0$\pm$0.5 \%). Here, we re-analyse the Herschel-PACS observations (fluxes from Mommert et al., 2012) including also the new SPIRE data, as well as updated fluxes from a revised reduction of the Spitzer-MIPS data. The thermal modelling of all these data (Fig.~\ref{huya}) gives size and albedo solutions very close to those obtained by Mommert et al (2012): D = 458$\pm$9.2 km, p$_V$=8.3$\pm$0.4 \%, and $\eta$=0.93$\pm$0.02 (Table~\ref{results}).\\
 Recently, Noll et al. (2012), analysing HST images of Huya, discovered the presence of a binary companion 1.4 mag fainter than the primary. With our size estimation and the given difference in magnitude of the companion, we estimate a diameter of 406$\pm$16 km for Huya, and of 213$\pm$30 km for the secondary. We cannot derive the density of the binary system because there is no mass estimation available yet.

\section{Classical (50000) Quaoar}

\begin{figure}
   \centering
\includegraphics[width=9.4cm,angle=0]{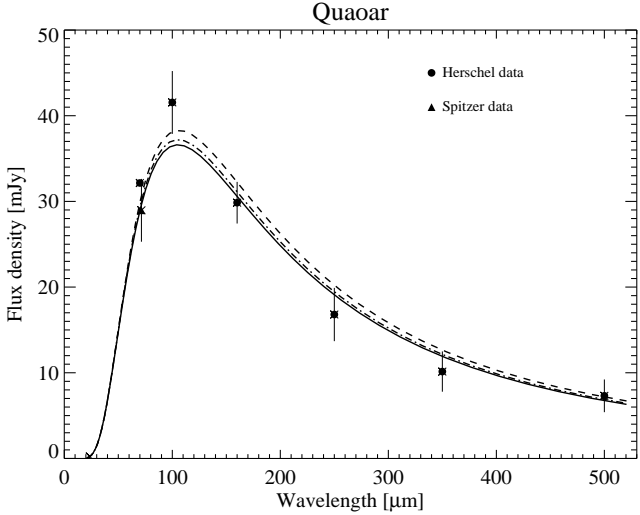}
      \caption{NEATM thermal model of Quaoar: model of the MIPS, PACS, and SPIRE data scaled to the heliocentric distance corresponding to the Herschel (solid line)  and to the Spitzer (dash--dotted line) observations; dashed line: model of the MIPS and PACS data scaled to the heliocentric distance corresponding to the Herschel observations.}
         \label{quaoar}
   \end{figure}

Quaoar is a TNO belonging to the classical population, and one of the largest transneptunians known to date.
Spectroscopic studies of Quaoar reveal that it is a very intriguing object, with a very red spectrum in the visible region (visual slope of $\sim$ 27 \%/(10$^3$ \AA), Fornasier et al. 2004a, Alvarez et al. 2008), and the presence of several absorption bands in the near infrared region attributed to water ice in the amorphous and cristalline state and to methane and ethane ices (Jewitt \& Luu, 2004; Schaller \& Brown, 2007; DalleOre et al. 2009). \\
Its rotational period was initially estimated to be approximately 17.68 hours, from a double-peaked
lightcurve (Ortiz et al. 2003), but recently the single peak solution with a rotational period of 8.839$\pm$0.003h was considered most realiable (Thirouin et al., 2010, Rabinowitz et al. 2007).
Quaoar size estimation was reported for the very first time by Brown \& Trujillo (2004) from direct imaging with the HST telescope. They derived a size of 1260$\pm$190 km, and a geometric albedo of 9.2$^{+3.6}_{-2.3}$ \%. Later size determination from thermal infrared measurements with the Spitzer Space Telescope have led to a smaller Quaoar size: Stansberry et al. (2008) give a diameter of  844$^{+207}_{-190}$ km, and Brucker et al. (2009) D = 908$^{+112}_{-118}$ km.    

Here we report new measurements with the Herschel Space Observatory (Table~\ref{fluxHerschel}) together with revised Spitzer-MIPS fluxes (Table~\ref{fluxSpitzer}). The SPIRE observations have been executed around 11 months after the PACS ones (Table~\ref{tabobs}). Quaoar is the only target for which we have a positive detection at 500 $\mu$m, even if the flux is slightly below the 3$\sigma$ level. \\
The results of the NEATM model are reported in Table~\ref{results} and shown in Fig.~\ref{quaoar}. In the model, all the data acquired with Herschel were scaled to the observer distance ($\Delta$) corresponding to the PACS observations.
The NEATM model solution that best fits all the MIPS, PACS, and SPIRE observations gives D=1036.4$\pm$30.7 km, p$_V$=13.7$^{+1.1}_{-1.3}$ \%, and a relatively high beaming factor $\eta$=1.66$\pm$0.08 (Table~\ref{results}), indicating important thermal inertia effects. This $\eta$ value is still within the mean value derived by Vilenius et al. (2012) on a sample of 19 classical TNOs ( $\eta_{mean}$ = 1.47$\pm$0.43).

The Quaoar emissivity slightly decreases beyond 200 $\mu$m. Excluding the SPIRE data, which covers wavelength most sensitive to emissivity effects, the model that best fits the MIPS and PACS observations gives a higher diameter of 1074$\pm$38 km.

 \begin{figure}
   \centering
\includegraphics[width=6.4cm,angle=90]{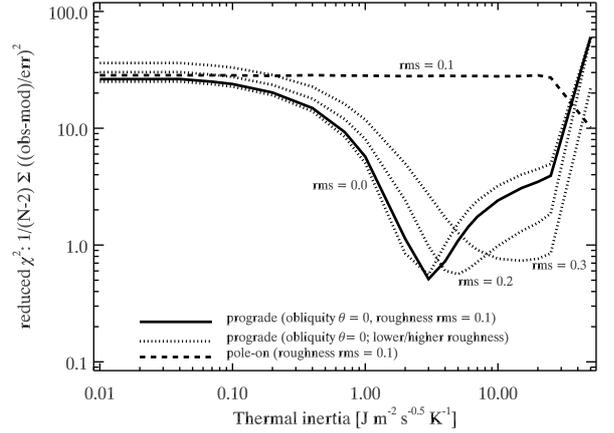}
      \caption{TPM analysis of Quaoar's thermal fluxes: $\chi^2$ test for different spin-vector orientations, and for different levels of roughness. $\theta$ is the obliquity. The pole-on solution can be excluded with high confidence.}
         \label{quaoar_tpm_chi2}
   \end{figure}
For Quaoar we also run a TPM model including all the available observations and considering a rotational period of  8.839$\pm$0.003h and a wavelength dependent emissivity which goes down to 0.8 at submm/mm wavelengths (M\"uller \& Lagerros 2002). The TPM model that best fits the data gives a strong indications that Quaoar is not pole-on, and that its surface must be smooth with a thermal inertia comprised between 2 and 4 in SI units (Fig.~\ref{quaoar_tpm_chi2}). Nevertheless it must be noted that TPM solutions having some surface roughness are possible with an associated larger thermal inertia value (5--20 in SI units).  Considering a thermal inertia value varing from 2 to 10, the TPM model gives an effective diameter comprised between 1015 and 1149 km, and a geometric albedo of 10-14\%.  \\
Considering that there is some degeneracy between roughness and thermal inertia for the TPM solutions, and that there are some emissivity effects for wavelengths $>$ 200 $\mu$m, our favoured solution for the Quaoar size and albedo is that derived from the NEATM analysis, excluding the SPIRE data (that are affected by emissivity effects).  
Our size estimation is in perfect agreement with the Quaoar size derived from occultations: Braga-Ribas et al. (2011, 2012) obtained a Quaoar's size ranging from  1020 km to 1212 km, and Braga-Ribas et al. (2013)  give a preferred solution for the effective diameter of 1111$\pm$5km from their Quaoar occultations campaigns. 

Quaoar has been found to be a binary system  with a smaller companion, Weywot, being 5.6$\pm$0.2 magnitudes fainter than the primary. The mass of the Quaoar/Weywot system was first determined to be of 1.6$\pm$0.3$\times$ 10$^{21}$ kg, and 1.65$\pm$0.16$\times$ 10$^{21}$ kg, by Fraser \& Brown (2010) and Vachier et al. (2012), respectively. Recently, Quaoar mass has been determined to be 1.4$\pm$0.1 $\times$ 10$^{21}$ kg from Fraser et al. (2013).
On the basis of the first mass evaluation and of a best guess of Quaoar size of 890$\pm$70 km, Fraser \& Brown (2010) derived a very high density value for Quaoar (4.2$\pm$1.3 g cm$^{-3}$), claiming that it is a rock body in Kuiper belt with little ice content. Using the same value of Quaoar size,  Fraser et al. (2013) found a bulk density ranging from 2.7 to 5.0 g cm$^{-3}$. \\
 With our size determination (D=1073.6$\pm$37.9 km), and considering recent refined estimation of the Quaoar/Weywot mass (Fraser et al. 2013), we derive a density of 2.18$^{+0.43}_{-0.36}$ g cm$^{-3}$, and, assuming the same albedo for the 2 bodies, a diameter of 1070$\pm$38 km for Quaoar, and 81$\pm$11 km for Weywot. The Quaoar density is considerably larger than that of pure water ice but it is not peculiar being similar to that of the dwarf planets Pluto and Haumea, and it indicates large amounts of rocky materials. Given the widespread presence of ices on the Quaoar surface, the comparatively large density implies a differentiated body.

\section{Hot Classical (55637) 2002 UX25}

\begin{figure}
   \centering
\includegraphics[width=9.4cm,angle=0]{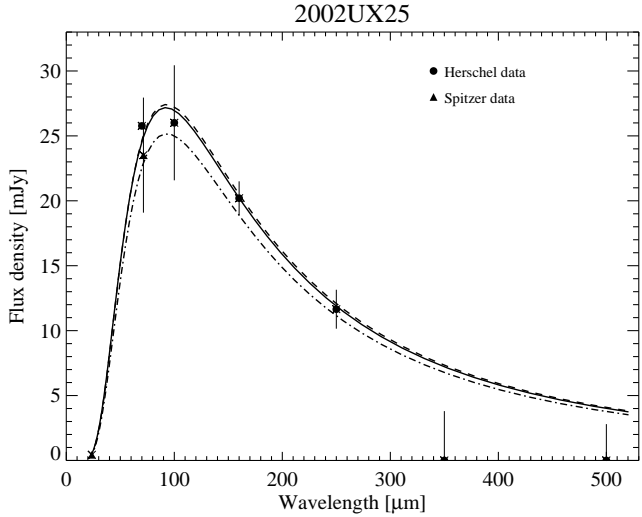}
      \caption{NEATM thermal model of 2002UX25: model of the MIPS, PACS, and SPIRE data scaled to the heliocentric distance corresponding to the Herschel (solid line)  and to the Spitzer (dash--dotted line) observations; dashed line: model of the MIPS and PACS data scaled to the heliocentric distance corresponding to the Herschel observations.}
         \label{2002UX25}
   \end{figure}

2002 UX25 is a hot classical TNO. Its spectrum is featureless and red in the visible-NIR region (Perna et al. 2010) but with a negative spectral slope in the 2.05-2.3 $\mu$m range, that makes it a possible candidate to have methanol-like compounds in its surface (Barucci et al., 2011). \\
Stansberry et al. (2008) derived a diameter of  680$^{+120}_{-110}$ km and a geometric albedo of 12$^{+5}_{-3}$ \% from Spitzer-MIPS observations. In this paper we present revised fluxes from Spitzer-MIPS (Table~\ref{fluxSpitzer}) together with new data from Herschel PACS and SPIRE (Table~\ref{fluxHerschel}) instruments. 2002 UX25 is clearly seen at 70, 100, 160, and 250 $\mu$m images, but it is undetected at 350 and 500 $\mu$m. Our best guess of its diameter and albedo is reported in Table~\ref{results}, and the values are similar, but with lower uncertainties, than those obtained by Stansberry et al. (2008). The NEATM models are presented in Fig~\ref{2002UX25}. The SPIRE data do not pratically affect the diameter and albedo estimation. 2002 UX25 shows an important decrease of the emissivity for wavelengths beyond $\sim$ 300 $\mu$m (Fig.~\ref{emi_all}).

Brown \& Suer (2007) reported the discovery of a satellite of 2002 UX25 2.5$\pm$0.2 mag fainter than the primary. Assuming a the same albedo for the primary and the secondary, with our size estimation we derive a diameter of 665$\pm$29 km for the primary and of 210$\pm$30 km for the secondary. Up to date, any estimation of the mass for  this binary system is available, so we cannot provide its density value. \\

\section{Resonant (84522) 2002 TC302}

\begin{figure}
   \centering
\includegraphics[width=9.4cm,angle=0]{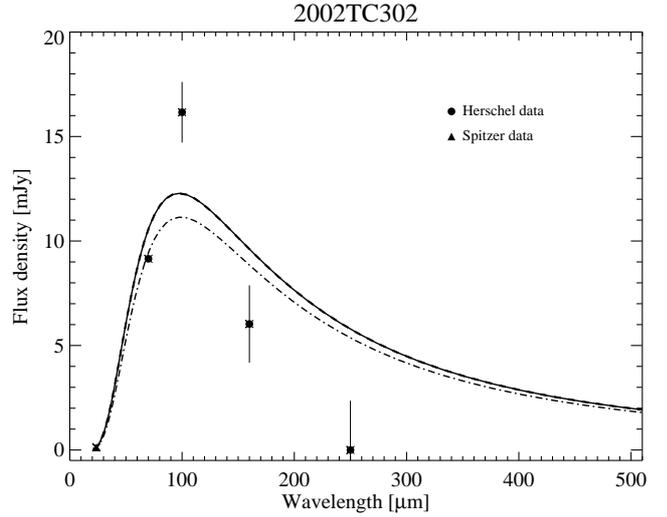}
      \caption{NEATM thermal model of 2002TC302: model of the MIPS, PACS, and SPIRE data scaled to the heliocentric distance corresponding to the Herschel (solid line)  and to the Spitzer (dash--dotted line) observations. The model of the MIPS and PACS data (dashed line), scaled to the heliocentric distance corresponding to the Herschel observations, is indistinguishable from the solid line.}
         \label{2002TC302}
   \end{figure}

2002 TC302 is a resonant object located in the 2:5 resonance with Neptune. It is a red object (Santos-Sanz et al., 2009), and its surface seems to be covered by some water ice (Barkume et al. 2008). Observations in the millimeter range gave a first estimation of 2002 TC302 diameter to be less than 1211 km with an albedo $>$ 5.1\% (Altenhoff et al., 2004; Grundy et al., 2005).  Stansberry et al. (2008) reported the detection of this body with MIPS in the two 24 and 70 $\mu$m  bands with SNR $<$ 5, and they estimated a diameter of 1150$^{+337}_{-325}$ km, a geometric albedo of 3.1$^{+2.9}_{-1.2}$, and $\eta$ = 2.3$\pm$0.5. Nevertheless, a re-analysis of the 2002 TC302 MIPS images shows that there was insufficient motion to allow a good sky subtraction and that the target happened to be very near a much brighter background object, making impossible the backgroud substraction at 71 $\mu$m. The MIPS revised 24 $\mu$m flux is reported in Table~\ref{fluxSpitzer}.\\
We observed this object with Herschel in 2010 and 2011, and it was detected with PACS but not with SPIRE (Table~\ref{fluxHerschel}).\\
The NEATM model that best fits the data (Fig.~\ref{2002TC302}) does not well reproduce the PACS observations, especially at 100$\mu$m.
Our best guess of its diameter is 584.1$^{+105.6}_{-88.0}$ km with an associated geometric albedo of 11.5$^{+4.7}_{-3.3}$ \%, revealing that this body is considerably smaller and brighter compared to the first estimations given by Stansberry et al. (2008).
Due to the lack of detections at longer wavelengths with SPIRE and to the poor NEATM fit, nothing more can be said about the emissivity properties of this resonant TNO.

\section{Plutino (90482) Orcus}

\begin{figure}
   \centering
\includegraphics[width=9.4cm,angle=0]{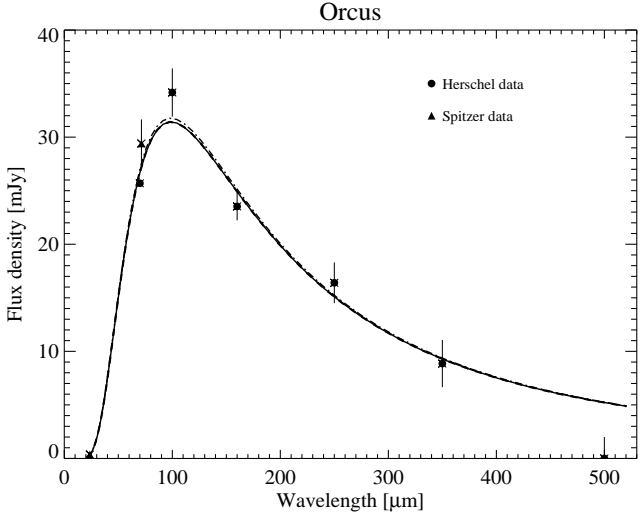}
      \caption{NEATM thermal model of Orcus:  model of the MIPS, PACS, and SPIRE data scaled to the heliocentric distance corresponding to the Herschel (solid line)  and to the Spitzer (dash--dotted line) observations. The model of the MIPS and PACS data (dashed line), scaled to the heliocentric distance corresponding to the Herschel observations, is indistinguishable from the solid line.}
         \label{orcus}
   \end{figure}

\begin{figure}
   \centering
\includegraphics[width=6.5cm,angle=90]{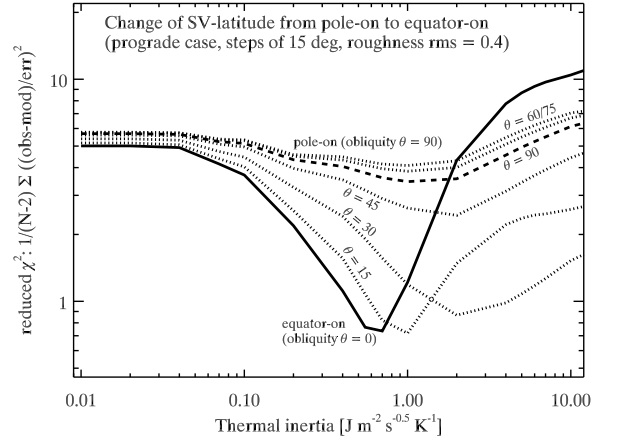}
      \caption{TPM analysis of Orcus' thermal fluxes: $\chi^2$ test for different spin-vector orientations. $\theta$ is the obliquity.}
         \label{orcus_inertia}
   \end{figure}

Orcus is a 1000 km-sized object (Stansberry et al. 2008), known to have a satellite, Vanth, and a surface rich in water ice in both amorphous and crystalline forms (Fornasier et al. 2004b; de Bergh et al., 2005; Trujillo et al., 2005; Barucci et al., 2008, Guilbert et al., 2009; Delsanti et al., 2010, DeMeo et al., 2010). It is one of the few binary TNOs in an orbit resonant with that of Neptune (it is a plutino populating the 3:2 resonance with Neptune). \\
A first analysis of the Orcus Herschel-SPIRE data together with Spitzer measurements was presented by Lim et al. (2010). They found an albedo of 0.25$\pm$0.03, a diameter of 867$\pm$57 km, and a beaming factor $\eta$ of 0.97$\pm$0.07. The SPIRE data have been re-analysed in the present paper with a revised instrument calibration, and with updated reduction and flux extraction procedures. This is why the SPIRE fluxes (Table~\ref{fluxHerschel}) are slightly different compared to those presented by Lim et al. (2010). In the meantime, we acquired additional observations of Orcus with PACS at 70-100-160 $\mu$m, and we got revised Spitzer MIPS fluxes. The data were modelled with both NEATM (Fig.~\ref{orcus}), and TPM thermal models, giving very similar results in terms of diameter and albedo (Table~\ref{results}), but with larger uncertainties for the TPM solutions: the diameter ranges from 915 to 1030 km, and the albedo is comprised between 20\% and 26\%. The TPM analysis is based on a 10.47 h rotation period, and it gives a diameter of 967.6$\pm$62 km and p$_V$= 22.9$\pm$3.0\% as the best solution. Our preferred solution is that given by the NEATM model: D = 958.4$\pm$22.9 km and  p$_V$= 23.1$^{+1.8}_{-1.1}$. \\
These values are slightly different from those derived by Lim et al. (2010), and closer to the ones got by Stansberry et al. (2008) and Brown et al. (2010) on the basis of the Spitzer data alone (Stansberry et al. (2008) obtained D=946$\pm$74 km, and p$_V$=0.197$\pm$0.034; Brown et al. (2010) got D=940$\pm$70 km and p$_V$=0.28$\pm$0.04). \\
The NEATM model gives an $\eta$ value of 0.97$^{+0.05}_{-0.02}$. This value is fully consistent with the mean $\eta$ value determined for 18 plutinos (1.11$_{-0.19}^{+0.18)}$) from combined Herschel/Spitzer data (Mommert et al. 2012). \\
In a general sense, $\eta$ values close to 1 indicate a thermal regime close to slow rotator, i.e. a low thermal inertia.
Indeed, the TPM analysis indicates that Orcus has a low thermal inertia value in the range 0.5-2.0 J m$^{-2}$ s$^{-0.5}$ K$^{-1}$, and that during the observations with Herschel and Spitzer Orcus was very likely close to an equator-on geometry (Fig.~\ref{orcus_inertia}), with an obliquity less than 30 $^o$, and not in a pole-on geometry that might have been expected as the orbit of Orcus's moon Vanth is essentially circular and seen pole-on (Brown et al. 2010). The low amplitude of Orcus's rotational lightcurve (0.03-0.04 mag, Thirouin et al. 2010) is instead in favour of a pole-on solution, or of a nearly spherical shape of Orcus.

Brown et al. (2010) determined a mass of (6.32$\pm$0.05)$\times$ 10$^{20}$ kg for the Orcus/Vanth system, a difference in magnitude of  2.54$\pm$0.01 mag between the primary and the secondary, a density of 1.4$\pm$0.3 g cm$^{-3}$, and individual diameters of 900 km and 280 km for Orcus and Vanth, respectively, assuming a similar albedo value for the 2 bodies. \\
Using a diameter of 958.4 km, which is our best guess of the Orcus/Vanth system size, we get a density of 1.53$^{+0.15}_{-0.13}$ g/cm$^3$,  with individual diameters of 917$\pm$25 km and 276$\pm$17 km for the primary and secondary, respectively.

\section{Classical (120347) Salacia}

\begin{figure}
   \centering
\includegraphics[width=9.4cm,angle=0]{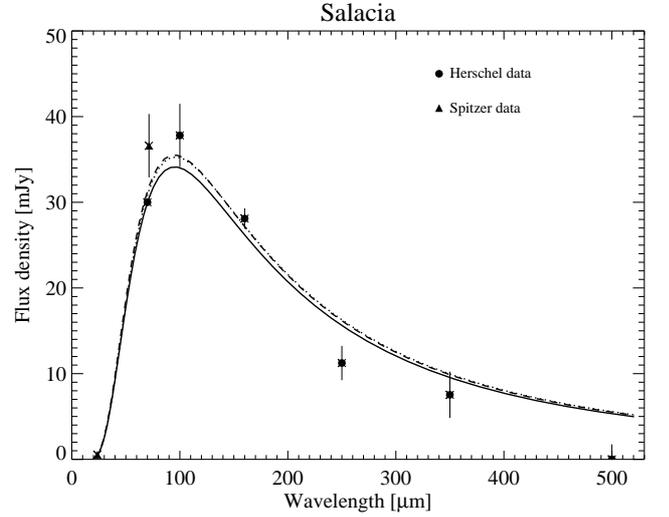}
      \caption{NEATM thermal model of Salacia: model of the MIPS, PACS, and SPIRE data scaled to the heliocentric distance corresponding to the Herschel (solid line)  and to the Spitzer (dashed line) observations; dotted line: model of the MIPS and PACS data scaled to the heliocentric distance corresponding to the Herschel observations.}
\label{Salacia}
\end{figure}

Salacia, formely 2004 SB60, is a classical binary TNO. Spitzer and Herschel-PACS observations of Salacia were reported by Vilenius et al. (2012), who found a diameter 901$\pm$45 km, a geometric albedo of 4.4$\pm$0.4 \%,  and a beaming factor of  1.16$\pm$0.03. Stansberry et al. 2012, from Spitzer only observations, obtain a diameter  of 954$\pm$109 km, and p$_V$= 3.57$^{+1.03}_{-0.72}$. \\
In this paper we present new data from SPIRE (Table~\ref{results}). The diameter resulting from the NEATM model of the MIPS, PACS, and SPIRE data is 874$\pm$32 km, slightly smaller than obtained when excluding the SPIRE data (901$\pm$45) but still within the error bars. In both cases, the SPIRE fluxes are lower than predicted by the NEATM best fit model, indicating a lower emissivity of Salacia at larger wavelengths (see Fig.~\ref{emi_all}). 

Salacia is a binary system and the secondary, Actea, is 2.37$\pm$0.06 magnitudes fainter. The mass of the system was first determined by Grundy et al. (2011) to be (4.66$\pm$0.22)$\times$10$^{20}$ kg, and recently refined by Stansberry et al. (2012) to be (4.38$\pm$0.16)$\times$10$^{20}$ kg. With this last value of the mass, and considering the diameter of the system (Table~\ref{results}) derived from MIPS and PACS observations, we obtain a density of the system of 1.29$^{+0.29}_{-0.23}$ g/cm$^3$. Assuming the same albedo for the primary and the secondary, we derive a diameter of 854$\pm$45 km for Salacia, and 286$\pm$24 km for Actea. Using the MIPS data alone, Stansberry et al. (2012) derived a density of  1.16$^{+0.59}_{-0.36}$ g/cm$^3$, and diameters of 905$\pm$103 km and 303$\pm$35 km for Salacia and Actea, respectively. This binary system  is very similar to the Orcus/Vanth system concerning the bodies individual size, but it has a smaller density value, indicating a higher percentage of water ice in its interior compared to Orcus. The surface properties appear indeed very different, as Orcus spectra show abundant water ice bands (Fornasier et al. 2004b, Barucci et al. 2008), while the near-infrared spectrum of Salacia is essentially featureless with an estimated aboundance of water ice less than 5\% (Schaller \& Brown, 2008).

\section{The dwarf planet (136108) Haumea}

The dwarf planet 136108 Haumea is one of the most intriguing bodies among the population of Kuiper-Belt objects. It is the largest member of a dynamical family characterised by water-ice rich surfaces (Ragozzine \& Brown, 2007; Trujillo et al. 2007), it has a fast rotational period, two close satellites (Brown et al. 2007), and an oblong shape (Rabinowitz et al. 2006). Its spectrum is caracterised by a neutral slope in the visible and by deep absorption bands in the NIR region associated to water ice both in the crystalline and amorphous form (Merlin et al., 2007, Pinilla-Alonso et al., 2009).
Time-resolved optical photometry of Haumea has revealed evidence for a localised surface feature both redder and darker than the surrounding material (Lacerda et al. 2008); this feature is often
referred to as the dark red spot. Lacerda (2009) found visible and near-infrared photometric colour variations across the surface of Haumea, variations which are spatially correlated with the dark spot. Nevertheless, Pinilla-Alonso et al. (2009) suggest that the Haumea surface must be very homogeneous as no variations larger than the measurement uncertainties were found in the spectra acquired at different rotational phases. \\
The Haumea size was first determined with lightcurve studies by Rabinowitz et al. (2006), who found D = 1350$\pm$100 km and p$_V$=0.65$\pm$0.06. Stansberry et al. (2008), using Spitzer MIPS observations, reported a diameter of 1150$^{+250}_{-100}$ km and an albedo of 84$^{+10}_{-20}$ \%.
The first observations of Haumea obtained within the {\it TNOs are Cool programme} were presented in Lellouch et al. (2010). They observed the dwarf planet with PACS at 100 and 160 $\mu$m covering altmost the full Haumea rotational period. Combining the PACS observations with the MIPS ones, they found a diameter comprised between $\sim$ 1210-1300 km for NEATM solutions with fixed $\eta$ value ($\eta$ fixed to 1 and 1.2), and a surface caracterised by a low thermal inertia, and probably covered by fine regolith.\\
Lellouch et al. (2010) obtained also the Haumea thermal lightcurve. They found a large amplitude indicative of a large a/b ratio ($\sim$ 1.3), and of a moderately low thermal inertia ($\eta$ $<$ 1.15-1.35).

In this paper we present new fluxes for both SPIRE and PACS instruments (Table~\ref{tabobs}), covering the 6 bands from 70 to 500 $\mu$m (Table~\ref{fluxHerschel}). Haumea was observed with PACS on 20 June 2010 for a full rotational period in the 100 and 160 $\mu$m band, and on 21 June 2010 with the 70 and 100 $\mu$m band covering just a small fraction of the rotational period.
The analysis of Haumea lightcurve will be presented in a separate paper (Santos-Sanz et al., in preparation). In Table~\ref{fluxHerschel} we report the obtained 70 $\mu$m flux and the mean fluxes from the 100 and 160 $\mu$m lightcurve observations.  With Spitzer-MIPS, Haumea was not detected at 24 $\mu$m but only at 70 $\mu$m (Table~\ref{fluxSpitzer}). Even without the 24 $\mu$m flux density, the NEATM with free $\eta$ run over all the MIPS, PACS and SPITZER fluxes gives a reasonable value of the beaming factor, 0.95$^{+0.33}_{-0.26}$, consistent with the upper limits derived by  Lellouch et al. (2010) from the analysis of Haumea thermal lightcurve.
Our NEATM solution gives a size comprised between 1182 and 1308 km, and a geometric albedo value of 80$^{+6}_{-10}$\% (Table~\ref{results}). No emissivity effects below 500 $\mu$m are seen for Haumea (Fig.~\ref{emi_all}).

\begin{figure}
   \centering
\includegraphics[width=9.4cm,angle=0]{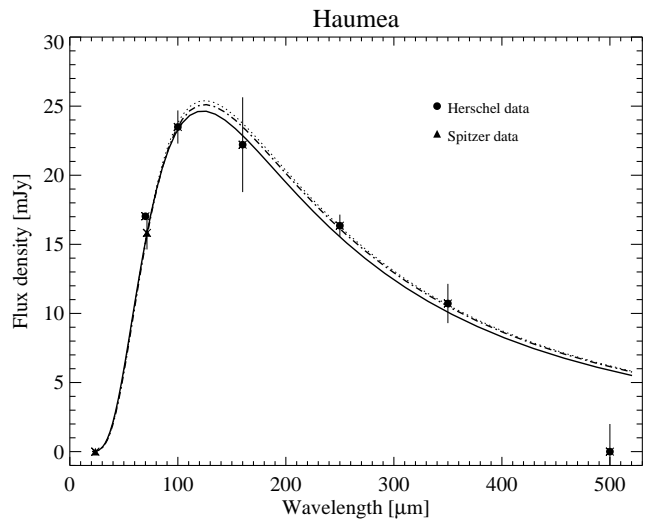}
      \caption{NEATM thermal model of Haumea: model of the MIPS, PACS, and SPIRE data scaled to the heliocentric distance corresponding to the Herschel (solid line)  and to the Spitzer (dash--dotted line) observations; dotted line: model of the MIPS and PACS data scaled to the heliocentric distance corresponding to the Herschel observations.}
         \label{Haumea}
   \end{figure}

\section{Discussion}

The TNOs and Centaurs here presented may be considered representative of various kind of minor bodies in the outer solar system. Indeed, they belong to different dynamical classes (Centaurs, plutinos, classicals TNOs), with diameters ranging from $\sim$ 220 to 1300 km, and have different surface properties: low to high albedo (4--80\%) objects, some have a surface composition dominated by water ice (like Orcus or Haumea), or a combination of water and methane ices (Quaoar), or other non-ice components (amorphous carbon, silicates, tholins) (Barucci et al. 2011). \\
From the compilation of Lamy et al. (2004), geometric albedos of cometary
nuclei are in the 2-6 \% range. The albedo of Chariklo is measured
to be right in this range, whereas that of Chiron is found to be much higher (Table~\ref{results}). For the nucleus of the giant comet C/1995 O1 (Hale-Bopp), recent
estimations based upon Herschel observations give a diameter of 94$\pm$15 km
and an albedo of 5.5$\pm$1.9 \% from the preliminary analysis of Kidger et
al. (2012), and a diameter of 74 $\pm$6 km and an albedo of 8.1 $\pm$ 0.9 \% from the study of Szab\'o et al.  (2012). This places the albedo of comet Hale-Bopp as intermediate between that of standard comets and Chiron, which indeed shows a peculiar high albedo.
Clearly, the size, and therefore the gravity, of active icy objects is an
important paremeter to consider when trying to understand the evolution of
their surface properties.

The combined thermal measurements of different instruments covering the 24-500 $\mu$m range
allow us not only to derive size and albedo of the investigated targets but also to study their thermal properties.
Indeed, the thermal emission  depends not only on the body's size and albedo, but also on surface properties such as thermal inertia, spin state, surface roughness, and emissivity.\\
The beaming parameter $\eta$ used in the NEATM models represents empirically the combined effects of thermal inertia and surface roughness. The $\eta$ values here obtained range from 0.9 to 1.2 for all the targets except for Quaoar, that shows a higher $\eta$ value of about 1.7.  Our findings on $\eta$ agree
with results based on a sample of 18 plutinos, 19 classical, and 15 scattered disk/ detached objects (Mommert et al. 2012, Vilenius et al. 2012,
Santos-Sanz et al. 2012). These sample results led to average values of $\eta$ of 1.11$^{+0.18}_{-0.19}$, 1.47$\pm$0.43, and 1.14$\pm$0.15 for
the three families, respectively. Quaoar's relatively high $\eta$ value is not anymore an outlier in the context of Classical objects. Note also
that the comparison of the TPM-inferred best fit thermal inertias of Quaoar and Orcus (2-4 and 0.5-2.0 J m$^{-2}$ s$^{-0.5}$ K$^{-1}$, respectively, for the best fit solutions), two objects observed at similar heliocentric distances, confirms the relationship between $\eta$ and thermal inertia.  

The SPIRE measurements are particularly valuable to assess the long-wavelength behaviour of the SED, related to the surface  emissivity. In our thermal models, $\epsilon$ is assumed equal to 0.9 and grey at all wavelengths.
The wavelength-dependent emissivity $\epsilon(\lambda)$ can be determined by comparing the observed fluxes to the NEATM
best fit. Figure~\ref{emi_all} shows the ratio between the
observed fluxes obtained with Herschel and Spitzer (and WISE for Chariklo) and the NEATM best fit, as a function of wavelength, for all targets except 2002 TC302, which was not detected with SPIRE. \\
All the investigated bodies show a significant decrease in $\epsilon$ towards the longest wavelengths. For most objects, the decrease starts in the SPIRE range, i.e. longwards of 200 $\mu$m, but the onset of emissivity effects varies from object to object. It starts at 250 $\mu$m for Quaoar and Salacia, 350 $\mu$m for 2002 UX25 and Chariklo, and is visible
only at 500 $\mu$m on Huya, Orcus, and Haumea. Chiron shows the strongest emissivity effects, with a decrease for wavelengths $\ge$ 100 $\mu$m (Fig.~\ref{emi_all}). None of the objects was detected at 500 $\mu$m, except Quaoar for which a $\sim 2\sigma$ flux detection is seen. \\
Similar effects were observed in the past for main belt asteroids, icy satellites and comets, from observations at mm/sub-mm wavelengths (Redman et al. 1992, 1998; M\"uller \& Lagerros 1998; Mathews et al. 1990; Fernandez 2002; Boissier et al. 2011). M\"uller (2002) found a general drop of the emissivity  beyond 150 $\mu$m for the asteroids observed with ISO telescope. Asteroid 4 Vesta shows a low emissivity value of 0.6-0.75  at millimetre and sub-mm wavelengths (M\"uller \& Barnes 2007, Leyrat et al. 2012). The decrease in emissivity with $\lambda$ on asteroid surfaces is confirmed also by Rosetta/MIRO observations of 2867 Steins, with emissivities determined to be 0.6--0.7 and 0.85--0.9 at wavelengths of 0.53 and 1.6 mm, respectively (Gulkis et al. 2010). On the other hand, 21 Lutetia, also observed by MIRO during the July 2010  Rosetta fly-by, shows a relatively high submm emissivity (0.95), and surface thermal properties similar to Apollo lunar regolith (Gulkis et al. 2012).\\
 As emphasized by Gulkis et al. 2010, caution must be exercised when inter-comparing all these emissivity values.
In most of the above papers, as well as in the current work, ``emissivity'' is a effective quantity derived from ratioing the observed fluxes
to model predictions that assume that thermal radiation is emitted from the surface itself. (In the work of Redman et al. (1992) it further assumes that
the object is in a non-rotating state). In contrast, the emissivities derived from Rosetta / MIRO refer to a model including subsurface emission and surface
Fresnel reflection. \\
The submm range seems to be the region where the emissivity reaches its minimum value (Mathews et al., 1990; Redman et al. 1992; M\"uller \& Lagerros 1998; Gulkis et al. 2010).
 For asteroids, the $\epsilon(\lambda)$ behaviour was attributed to the presence of a dusty and porous regolith in which backscattering within the subsurface
reduces the outgoing emission. Redman et al. (1992) state that scattering by grains smaller than 100 $\mu$m within the regolith
can reduce the emissivity in the sub-mm region without affecting it at centimeter wavelengths.  Gulkis et al. (2010) also
mention scattering as the cause of subdued sub-mm emissivity but note that the strong emissivity variation between 0.53 and 1.6 mm is difficult
to understand. Scattering is also invoked to account for the very low brightness temperature of Ganymede at cm and radar wavelengths and its
anomalous radar reflectivity (Muhleman et al. 1991, Ostro \& Shoemaker 1990).    

For Centaurs and TNOs, the lower emissivity we infer at $\lambda >$ 250 $\mu$m may be attributed to the fact that the sub-mm thermal flux arises from sub-surface layers that are, on the dayside, colder than the surface itself. The same explanation has been proposed by Boissier et al. (2011) for the thermal emission from the nucleus of comet 8P/Tuttle. For definitiveness, we consider the cases of Orcus and Quaoar,
with nominal thermal inertia of 0.7 and 3 J m$^{-2}$ s$^{-0.5}$ K$^{-1}$ (values from the best TPM model fits, see Figs.~\ref{quaoar_tpm_chi2} and~\ref{orcus_inertia}.
Given their heliocentric distance, albedo and rotation rates (10.5 h and 8.9 h, respectively), these objects have thermal parameter $\Theta$
(as defined by Spencer et al. 1989, and representing the ratio of the timescale for radiation from the subsurface
to the diurnal timescale) equal to $\sim$1 and $\sim$4, respectively. These small values imply a significant temperature gradient
within the subsurface. The characteristic scale of the temperature gradient is given by the thermal skin depth,  $l_s$ = $\sqrt{k/\rho c
\omega}$ = $\Gamma/(\rho c \sqrt \omega)$, where $\omega$ = 2$\pi/\tau$, $\rho$ is
the density, $c$ is the heat capacity and
$k$ is the thermal conductivity. Using a typical $\rho$ = 930 kg~m$^{-3}$
and $c$ = 350 J~kg$^{-1}$~K$^{-1}$ for H$_2$O ice at $\sim$40 K, we obtain $l_s$ = 0.02 cm for Orcus and 0.06 cm for Quaoar.
These numbers are  similar to the wavelengths of the SPIRE range. \\
As thermal radiation is expected to originate
from layer depths at least several times larger than the wavelength, this implies that the SPIRE radiation likely probes
below the skin depth and thus ``sees" the diurnally-averaged ``deep" temperature. For  $\Theta$= 1 and 4,
the latter is typically 26 - 13 \% smaller that the maximum dayside temperature (e.g. Fig. 2 of Spencer et al. 1989),
which (in the Rayleigh-Jeans regime approximately valid of our observations) is consistent with comparable departures
from unit emissivity (i.e. emissivities of 0.75-0.85). We note however that in the framework of this interpretation, the steady decrease of the emissivity with wavelength implies that the shortest wavelengths (i.e. the PACS range)
do not probe the deep temperature below the diurnal wave, meaning that the surface cannot be too transparent. \\
Let us first consider surfaces covered by water ice. Matzler (1998) reviewed the microwave properties for ice and snow at 100-270 K. Extrapolating the
results (his Fig. 2) to $\sim$50 K indicates an absorption coefficient for pure ice of about 40 m$^{-1}$ at 300 $\mu$m
and varying as 1/$\lambda^2$. This gives an absorption depth of 2.5 cm at 300 $\mu$m, extrapolating to 0.25 cm at 100 $\mu$m. As this is several times larger than  $l_s$, this would imply that
the deep subsurface is probed even at 100 $\mu$m, in contradiction with the emissivity curve we infer.
Rather, the emissivity decrease we observe longwards of 200 $\mu$m implies absorption coefficients
about 10 to 20 times stronger than estimated from Matzler (1998). The same conclusion holds for surfaces covered by
methane ice: based on Stansberry et al. (1996), methane ice has an absorption coefficient of $\sim$2 cm$^{-1}$ at 100 $\mu$m,
which would imply an emission depth at least 10 times deeper than the thermal skin depth, contradicting our
observed emissivity behaviour. In both cases, we conclude that the surfaces have absorption properties at submm
wavelengths well enhanced over the pure ice behaviour. Impurities within the ices may be responsible for this enhanced absorption.

\section{Conclusions}

In this paper we have derived the size, albedo, and surface properties, including thermal inertia and surface emissivity, for nine bright TNOs and Centaurs observed by both the PACS and SPIRE multiband photometers onboard the Herschel space telescope: the dwarf planet Haumea, six TNOs (Huya, Orcus, Quaoar, Salacia, 2002 UX25, and 2002 TC302), and two Centaurs (Chiron and Chariklo). To derive the physical and thermal properties of these bodies, we ran the NEATM thermal model on the new data obtained from the Herschel Space Observatory over six bands (centred at 70, 100, 160, 250, 350, and 500 $\mu$m) combined with the revised fluxes from Spitzer-MIPS observations at 23.7 and 71.4 $\mu$m. For the Centaurs Chiron and Chariklo, and for the 1000 km sized Orcus and Quaoar, we also ran a thermophysical model with both constant and wavelength dependent emissivity to better constrain their thermal properties.\\
 Our main results are the following:
\begin{itemize}
\item For the Centaur Chiron, the NEATM and the TPM models give a similar albedo and size estimation. Our best estimation of Chiron diameter is 218$\pm$20 km, with an albedo value of 16$\pm$3\%, considerably higher than the values published before, and related to the bright
nucleus magnitude estimated close in time to the Herschel observations.\\
Chiron shows an important decrease in its emissivity for wavelengths $>$ 100 $\mu$m. Indeed, Chiron showed a considerable increase of its brightness after the 1999 perihelion passage, maintaining a bright magnitude over the past seven years. A re-analysis of visual images taken with the VLT telescope has shown that a faint coma was present in 2007-2008, while no coma was detected within our detection limit in the new visual images obtained on December 2011, or in the PACS 70 and 100 $\mu$m bands images. The high brightness over the past seven years may result from resurfacing following an activity outburst.
We derive an upper limit of 6--45 kg s$^{-1}$ for the dust production rate from the PACS images. These upper limits are 1.3-10 times higher than those derived from the optical images relative to the 2007-2008 observations.

\item  Centaur Chariklo:  our best estimation of its size comes from the TPM model with emissivity varying with wavelength in a Vesta-like manner. We find a diameter of 248$\pm$18 km and a very dark surface (2.5$ < p_V < $4.5\%), confirming that it is the largest Centaur known so far. Our TPM model also excludes the pole-on solution at the time of Herschel observations. We analysed visual images acquired in 2007-2008 and the Herschel PACS observations to constrain cometary activity and dust production rate. No coma was detected in these images, and the more stringent upper limit for the dust production rate (0.16--4 kg s$^{-1}$) was derived from the visual images.

\item We derived accurate size estimates for the $\sim$1000 km sized binary systems Quaoar and Orcus, using both NEATM and TPM models and assuming the same albedo for the primary and its satellite:  the diameter of Quaoar is 1070$\pm$38 km; that of its satellite, Weywot, is 81$\pm$11 km; and their geometric albedo is 12.7$\pm$1 \%. For the Orcus/Vanth system we estimate a diameter of 917$\pm$25 km and 276$\pm$17 km for the primary and the secondary, respectively, and a geometric albedo of 23.1$^{+1.8}_{-1.1}$ \%, higher than the plutinos' mean p$_V$ value (Mommert et al. 2012). From TPM analysis, both bodies have smooth/low roughness surfaces with thermal inertia values of 2-10 and 0.5-2.0 in SI units for Quaoar and Orcus, respectively.

\item We also derived the bulk densities (and the individual size of the primary-secondary, assuming an identical albedo) for the binaries Quaoar/Weywot (2.18$^{+0.43}_{-0.36}$ g/cm$^3$), Orcus/Vanth (1.53$^{+0.15}_{-0.13}$ g/cm$^3$), and Salacia/Actea (1.29$^{+0.29}_{-0.23}$ g/cm$^3$), using the available mass estimations and the size values derived by our observations with the Herschel telescope. In particular the Quaoar density is very different from the first estimations given by Fraser \& Brown (2010, i.e. 4.2$\pm$1.3 g cm$^{-3}$). The Quaoar density value reported here is close to that of the dwarf planets Pluto and Haumea, and it implies a high content of refractory materials mixed with ices.

\item  The dwarf planet Haumea is the largest body in our sample: we derived a diameter of 1240$^{+69}_{-58}$ km by the NEATM model of the Spitzer and Herschel fluxes, in line with previous estimations, and we confirmed the high geometric albedo value (80$^{+6}_{-10}$ \%).

\item Most targets show a significant decrease in their spectral emissivity longwards of $\sim$250 $\mu$m (but even at shorter wavelengths for the Centaur Chiron) and especially at 500 $\mu$m. The lower emissivity at $\lambda >$ 250 $\mu$m is attributed to the fact that the sub-mm thermal flux arises from sub-surface layers that are, on the dayside, colder than the surface itself. The steady decrease in the emissivity with wavelength implies that SPIRE radiation most likely probes below the skin depth, and thus is representative of the diurnally-averaged deep temperature, while PACS fluxes in the shorter wavelengths do not probe the deep temperature below the diurnal wave, indicating that the surface is not too transparent.
Comparing the decrease in emissivity with wavelength with the absorption depth derived from  microwave properties for water and methane ices, we conclude that the surfaces of the investigated Centaurs and TNOs have absorption coefficients at sub-mm wavelengths well enhanced over the pure ice behaviour. Impurities within the ices may be responsible for this enhanced absorption.

\end{itemize}

\begin{acknowledgements}

Herschel is an ESA space observatory with science instruments provided by European led Principal Investigator consortia and with important participation from NASA. Herschel data presented in this work were processed using HIPE, a joint development by the Herschel Science Ground Segment Consortium, consisting of ESA, the NASA Herschel Science Center, and the HIFI, PACS, and SPIRE consortia. This project was supported by the French Planetology National Programme (INSU-PNP), by the Hungarian Research Fund (OTKA) grant K104607, by the PECS programme of the European Space Agency (ESA), and the Hungarian Space Office (contract \#98073), by the Bolyai Research Fellowship of the Hungarian Academy of Sciences, and by the German {\it DLR} project number 50 OR 1108.

\end{acknowledgements}

\end{document}